\documentclass[fleqn,usenatbib]{mnras}

\usepackage{newtxtext,newtxmath}

\usepackage[T1]{fontenc}
\usepackage{ae,aecompl}

\usepackage{graphicx}	
\usepackage{amsmath}	
\usepackage{eucal}
\usepackage{enumitem}
\usepackage{lineno}
\usepackage{comment}
\usepackage{tikz}
\usepackage{gensymb}
\usepackage{newtxtext,newtxmath}
\usepackage{setspace}


\newcommand{\lsim}{\protect\raisebox{-0.8ex}{$\:\stackrel{\textstyle <}{\sim}\:$}} 
\newcommand{\gsim}{\protect\raisebox{-0.8ex}{$\:\stackrel{\textstyle >}{\sim}\:$}}

\newcommand\dbar[1]{
    \setbox0=\hbox{$\overline{#1}$}
    \ht0=\dimexpr\ht0-.25ex\relax
    \overline{\copy0}
}

\newcommand{\vpm}[3]{
    {#1}^{+ {\mkern+1mu} {#2}}_{{\mkern-1.5mu} -{#3}}
}

\defcitealias{2018MNRAS.478.1866H}{HE18}

\title[Quasar illumination and Ly$\alpha$ intensity mapping]{Prospects for constraining quasar ages with fiber spectrographs: \\ quasar-induced Ly$\alpha$ emission from the intergalactic medium}

\author[R. Hada et al.]{
Ryuichiro Hada,$^{1,2,3,4}$\thanks{E-mail: hada.555@osu.edu}
Masahiro Takada,$^{4,5}$ and 
Akio K. Inoue$^{6,7}$
\\
$^{1}$Center for Cosmology and AstroParticle Physics (CCAPP), The Ohio State University, Columbus, OH 43210, USA \\
$^{2}$Department of Physics, The Ohio State University, Columbus, OH 43210, USA \\
$^{3}$Department of Astronomy, The Ohio State University, Columbus, OH 43210, USA \\
$^{4}$Kavli Institute for the Physics and Mathematics of the Universe (WPI), UTIAS, The University of Tokyo, Kashiwa, Chiba 277-8583, Japan \\
$^5$Center for Data-Driven Discovery, Kavli IPMU (WPI), UTIAS, The University of Tokyo, Kashiwa, Chiba 277-8583, Japan \\
$^{6}$Department of Physics, School of Advanced Science and Engineering, Faculty of Science and Engineering, Waseda University, \\ 3-4-1, Okubo, Shinjuku, Tokyo 169-8555, Japan \\
$^{7}$Waseda Research Institute for Science and Engineering, Faculty of Science and Engineering, Waseda University, 3-4-1, Okubo, Shinjuku, Tokyo 169-8555, Japan \\
}

\date{Accepted XXX. Received YYY; in original form ZZZ}

\pubyear{\the\year}

\begin{document}
\label{firstpage}
\pagerange{\pageref{firstpage}--\pageref{lastpage}}
\maketitle

\begin{abstract} 
We present a theoretical framework for linking quasar properties, such as quasar age, to the surrounding Ly$\alpha$ emission intensity. In particular, we focus on a method for mapping the large-scale structure of Ly$\alpha$ emission intensity with galaxy spectra from wide-field spectroscopic surveys, e.g., the Subaru Prime Focus Spectrograph (PFS) or the Dark Energy Spectroscopic Instrument (DESI), and consider the quasar-induced Ly$\alpha$ emission from the intergalactic medium (IGM). To do this, we construct a theoretical model based on two physical processes: resonant scattering of quasar Ly$\alpha$ photons and fluorescence due to quasar ionizing photons, finding that the fluorescence contribution due to optically thick gas clouds is dominant. Taking into account the light cone effect and assuming a typical quasar spectrum, we calculate the fluorescence contribution to the spectrum stacked within each bin of the separation angle from the quasar as a function of quasar age. Furthermore, we compute the quasar-Ly$\alpha$ emission cross-correlation and its SNR for the planned PFS survey. The predicted signal can only account for $\sim10\%$ of the measurements indicated from the BOSS and eBOSS surveys in the outer region of $\gsim10\ \rm{cMpc}\ \rm{h}^{-1}$, and the predicted SNR is not sufficient to detect the quasar-induced contribution. However, we found that our model, combined with the contribution of star-forming galaxies, is not in conflict with these measurements. We discuss other possible contributions to the Ly$\alpha$ emission excess around quasars, the efficiency of using spectroscopic fibers, and the redshift dependence of our model.
\end{abstract}


\section{Introduction} \label{sec:intro}

While we have learned a lot of information about quasars from their emission line spectra \citep[e.g.,][]{1979RvMP...51..715D}, the ``life'' of quasars is still unknown because each quasar spectrum corresponds to a specific moment. In particular, the quasar lifetime is one of the key parameters that characterize quasars, and plays a critical role in understanding the activity of active galactic nuclei (AGN) or the growth of black holes. However, the current constraints are quite weak, lying in the range of $10^{6}{\rm -}10^{8} \ {\rm years}$. The large uncertainty is partly due to the fact that some quasar lifetime estimates rely on demographic methods, e.g., integrating quasar properties over cosmic time or comparing populations of objects at different redshifts. These approaches depend heavily on models that describe the (co)evolution of quasars, black holes, and galaxies, e.g., the quasar luminosity function or the black hole mass function. In addition, they can only set a limit on the {\it net} lifetime as the total amount of time that the accretion on to a supermassive black hole is luminous enough to be recognized as a quasar, rather than the {\it episodic} lifetime as the period of time of a single luminous phase \citep[see e.g.,][for a review]{2004cbhg.symp..169M}.
            
To probe the quasar activity more precisely, we then need to constrain its episodic lifetime (for an individual quasar, if possible) in a model-independent way. A promising way to estimate the episodic lifetime is to focus on the variation of the ionization state of the surrounding region. The radiation from a local quasar additionally ionizes the surrounding intergalactic medium (IGM) and then forms a characteristic region, which is called the {\it proximity effect (zone)} \citep{1988ApJ...327..570B}. Since the size of the ionized region depends strongly on the length of time the quasar remains active, we can constrain the quasar age (shorter than the episodic lifetime) without knowing the details of quasar physics. Observations of the proximity effect have been made by various studies and can be divided into two types: the line-of-sight and the transverse proximity effects \citep[e.g.,][for reviews of previous studies]{2019ApJ...882..165S,2021ApJ...917...38E}. 
    
The line-of-sight proximity effect can be seen in the quasar spectra themselves. Ionizing photons leaving the quasar earlier had ionized the surrounding IGM and then the Ly$\alpha$ forest absorption of the observed quasar spectrum should be reduced in the vicinity of the quasar \citep{1982MNRAS.198...91C,1988ApJ...327..570B}. Since all the photons that make up a quasar spectrum were emitted from the quasar at the same time, the propagation of the ionizing photons (the {\it ionization front}) must be considered in the quasar rest frame. Therefore, the size of the proximity zone on the spectrum is determined by the ionization timescale or the quasar age, whichever is shorter \citep[e.g., ][]{2000ApJ...542L..75C,2007MNRAS.374..493B}. However, at redshifts of $z\lsim5$, where a large fraction of the quasars ever observed lie, the measurements of the line-of-sight proximity zone only allow us to place a lower limit on the quasar age, because the ionization timescale is quite short, $\sim 10^{4}\ \rm{year}$.

The transverse proximity effect is detected by background sightlines (e.g., quasars or star-forming galaxies) passing close to a foreground quasar \citep{1989ApJ...336..550C,1991ApJ...377L..69D,2004ApJ...612..706A}. In this case, multiple spectra of background objects along different sightlines are less absorbed around the foreground quasar, and the variations in the spectra are caused by ionizing photons that, unlike the line-of-sight proximity effect, are unrelated to the currently observed quasar luminosity. This allows us to probe quasar ages much longer than the ionization timescale. Furthermore, at low redshifts where the hydrogen neutral fraction is very small, the ionization front propagates almost at the speed of light \citep[e.g.,][]{2006ApJ...648..922S}, which means that the size of the proximity zone depends only on the quasar age. On the other hand, since the difference in the optical depths for absorption systems inside and outside the proximity zone is characterized by the ratio of the quasar's ultraviolet (UV) flux to the UV background flux, we need to find background sightlines close enough to the foreground quasar to detect the proximity effect.

While we can study the effect of quasar radiation on the IGM through Ly$\alpha$ absorption lines in the spectra of background luminous objects by focusing on the proximity effect, Ly$\alpha$ emission from the surrounding IGM should be enhanced by the quasar illumination. The quasar-induced Ly$\alpha$ emission can be divided into two physical processes: resonant scattering and fluorescence \citep[e.g.,][]{Cantalupo2017}. When photons from a quasar pass through a nearby HI cloud, those observed as Ly$\alpha$ photons by HI atoms in the cloud (i.e., continuum photons slightly blueward of Ly$\alpha$ in the quasar rest frame) are resonantly scattered by the atoms ({\it resonant scattering}). In addition, Lyman continuum photons from the quasar, which form the proximity zone, promote photoionization in the HI cloud and produce additional Ly$\alpha$ photons through the subsequent recombination ({\it fluorescence}). These two processes would be observed by an observer in the direction in which the quasar photons initially propagated as a Ly$\alpha$ absorption line and a Lyman limit break, respectively, in the spectrum. This means that the excess of Ly$\alpha$ emission from HI clouds in the vicinity of quasars includes the sum of these two contributions. 

The detectability of Ly$\alpha$ emission from the IGM was first discussed theoretically in the context of imaging optically thick gas clouds in fluorescent Ly$\alpha$ emission that is induced by the UV background \citep{1987MNRAS.225P...1H,1996ApJ...468..462G}. Later studies then considered the effect of local quasars on the IGM Ly$\alpha$ emission by using a simple toy model \citep{2001ApJ...556...87H} or combining hydrodynamical cosmological simulations with a radiative transfer algorithm \citep{2005ApJ...628...61C,2010ApJ...708.1048K}, suggesting that the quasar ionizing photons enhance the fluorescent Ly$\alpha$ surface brightness of the surrounding IGM, in some high-density environments by a factor of $\sim 10^{2}$.
A series of narrow-band observations in the field of a quasar at $z\simeq3.2$ suggested for the first time that the Ly$\alpha$ companion cloud is fluorescently illuminated by its host quasar \citep[][]{1985ApJ...299L...1D,1987ApJ...317L...7H}, and since then a large number of observations have been reported investigating extended Ly$\alpha$ emission on scales of $\lsim 100\ {\rm pkpc}$ (physical kiloparsecs) \citep[e.g., ][for a review]{Cantalupo2017}. 

In analogy to the proximity effect, these results suggest that we can estimate quasar age from the size of regions where Ly$\alpha$ emission is enhanced by the quasar illumination. The size of this Ly$\alpha$-enhanced region is characterized by the light travel distance of the quasar age and is therefore much larger than the virial radius of the quasar host halo, i.e., $\gsim 1\ {\rm pMpc}$. There have been some efforts to explain various systems of quasars and their Ly$\alpha$ companions separated by relatively large distances with fluorescence enhanced by the quasars: a separated damped Ly$\alpha$ system \citep{2006ApJ...637...74A}, Ly$\alpha$ emitters (LAEs) \citep{2007ApJ...657..135C,2008ApJ...681..856R}, gas-rich proto-galactic clouds with very low star formation efficiencies (or {\it dark galaxies}) \citep{2012MNRAS.425.1992C,2018ApJ...859...53M}, or giant Ly$\alpha$ nebulae \citep[][]{2014Natur.506...63C,2014ApJ...786..106M,2015Sci...348..779H}. Observations of fluorescent LAEs around quasars actually set constraints on the quasar lifetime and emission opening angle by comparing the spatial distribution with models \citep{2013ApJ...775L...3T,2016ApJ...830..120B}. On the other hand, the contribution from resonant Ly$\alpha$ scattering of quasar continuum photons can only dominate in the diffuse, low-density IGM which is optically thin to ionizing photons, e.g., in the Ly$\alpha$ forest \citep[e.g.,][]{2010ApJ...708.1048K,2013ApJ...766...58H}. Therefore, we only need to consider the fluorescence process when quantifying the effect of quasar illumination on optically thick objects in the IGM, as in the previous works mentioned above, although the continuum scattering could be important in regions close enough to quasars that the gas is highly ionized by the quasar radiation \citep[e.g.,][]{2013ApJ...766...58H,2014ApJ...786..106M,2019MNRAS.482.3162A} or in the case where the neutral hydrogen number density is high enough to keep the gas neutral \citep{2014Natur.506...63C,Cantalupo2017}.

\begin{figure*}
  \begin{center}
   \includegraphics[width=0.9\linewidth]{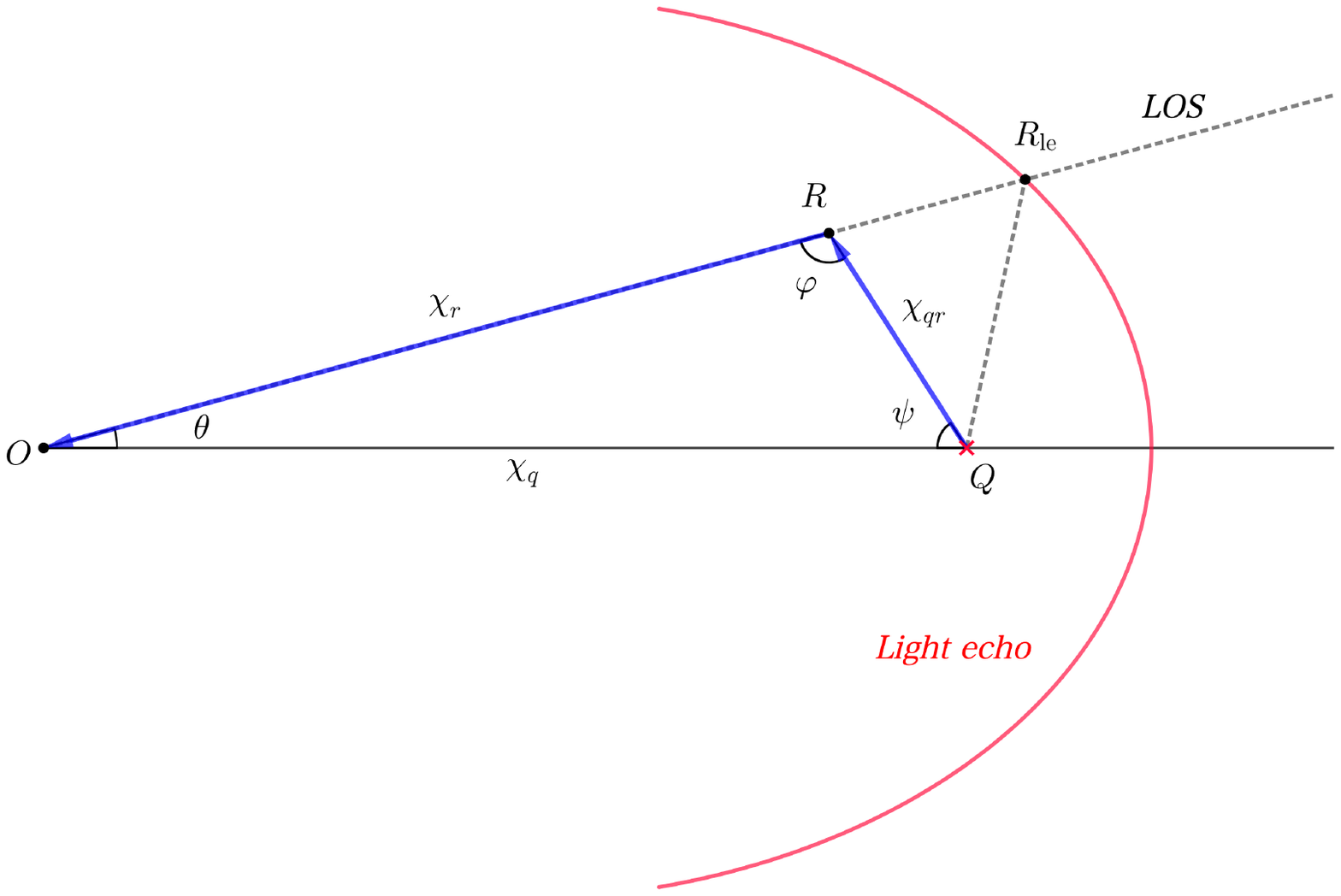}
  \end{center}
  \caption{\label{fig:schematic_LE} Schematic view of a light path for quasar-induced Ly$\alpha$ photons. Radiation emitted from a quasar of $Q$ at $t<t_{\rm age}$, where $t_{\rm age}$ is the quasar age, induces Ly$\alpha$ photons at the point of $R$ in the direction of $\theta$ with respect to the line of sight from an observer of $O$ to the quasar, and then the induced Ly$\alpha$ photons are observed ($Q \to R \to O$, blue solid line), simultaneously with photons that left the quasar at $t=t_{\rm age}$ and propagated directly to the observer ($Q \to O$). The red solid line shows the outermost region (quasar light echo) where Ly$\alpha$ photons are induced by quasar photons emitted at $t=0$.}
\end{figure*}

In this paper, we focus on a method for mapping the large-scale structure of Ly$\alpha$ emission intensity with galaxy spectra from wide-field spectroscopic surveys, and consider the contribution of Ly$\alpha$ emission driven by quasar radiation, with the goal of linking the Ly$\alpha$ intensity map to quasar properties. \citet{2016MNRAS.457.3541C,2018MNRAS.481.1320C} mapped the Ly$\alpha$ emission intensity distribution from luminous red galaxy (LRG) spectra from the Baryon Oscillation Spectroscopic Survey (BOSS) of the Sloan Digital Sky Survey III \citep[SDSS-III,][]{2011AJ....142...72E} by subtracting the best-fitting galaxy spectra and found a positive signal in the cross-correlation between the Ly$\alpha$ emission and BOSS quasars. Moreover, in combination with the cross-correlation between the Ly$\alpha$ emission and the Ly$\alpha$ forest sample constructed from BOSS quasar spectra, they concluded that a simple model where the Ly$\alpha$ emission intensity is proportional to the baryon density and the quasar flux can successfully explain both cross-correlation signals. However, \citet{2022ApJS..262...38L} recently revisited this conclusion while applying the same intensity mapping technique to the extended BOSS (eBOSS) data of SDSS-IV \citep{2017AJ....154...28B} and argued that a different model, where star-forming galaxies are responsible for the Ly$\alpha$ intensity distribution on scales of $\gsim 1\ {\rm cMpc}$, is preferred, since the quasar-based model requires an extremely high Ly$\alpha$ luminosity per quasar. 

We will soon obtain an even larger number of galaxy spectra with higher signal-to-noise ratios of individual spectra from the ongoing and upcoming wide-area spectroscopic surveys such as the Subaru Prime Focus Spectrograph \citep[PFS,][]{2014PASJ...66R...1T}\footnote{\url{https://pfs.ipmu.jp/}}, the Dark Energy Spectroscopic Instrument \citep[DESI,][]{2016arXiv161100036D}\footnote{\url{https://www.desi.lbl.gov/}}, Euclid \citep[][]{2011arXiv1110.3193L}\footnote{\url{https://www.esa.int/Science_Exploration/Space_Science/Euclid}}, and the Nancy Grace Roman Space Telescope \citep[][]{2015arXiv150303757S}\footnote{\url{https://roman.gsfc.nasa.gov}}. Motivated by these promising prospects, in this paper we theoretically model the boosted Ly$\alpha$ emission from the IGM illuminated by a local quasar, taking into account the effect of the light cone, and estimate its contribution to the spectrum stacked within each bin of the separation angle from the quasar. In particular, we consider both resonant scattering and fluorescence effects and compare the contributions from optically thin and thick gas clouds for each effect. Furthermore, we compare our prediction of the quasar-induced contribution to the Ly$\alpha$ emission excess with previous measurements or with that expected from star-forming galaxies clustered around quasars.

This paper is structured as follows: in Section~\ref{sec:light_cone}, we explain possible light paths for Ly$\alpha$ photons induced by quasar radiation and express them, including the quasar light echo, as a function of quasar age, redshift, and line-of-sight separation. We present some models for the neutral hydrogen column density distribution to describe the IGM, the typical quasar spectrum, and the clustering of star-forming galaxies in Section~\ref{sec:igm_qso}. We then consider the Ly$\alpha$ emissions driven by the quasar illumination in two different physical processes: resonant scattering (Section~\ref{sec:scattering}) and fluorescence (Section~\ref{sec:fluorescence}), and estimate the contribution to the stacked spectrum (Section~\ref{sec:quasar-induced_Lya}). In Section~\ref{sec:cross-corr}, we calculate the cross-correlation between quasar and Ly$\alpha$ emission and evaluate the signal-to-noise ratio for the PFS survey. We discuss the limitation, efficiency, and redshift dependence in Section~\ref{sec:discussion} and summarize the results and future prospects in Section~\ref{sec:conclusion}. In the following, we adopt the cosmological model that is consistent with \citet{2016A&A...594A..13P}.

\begin{figure}
  \begin{center}
   \includegraphics[width=\linewidth]{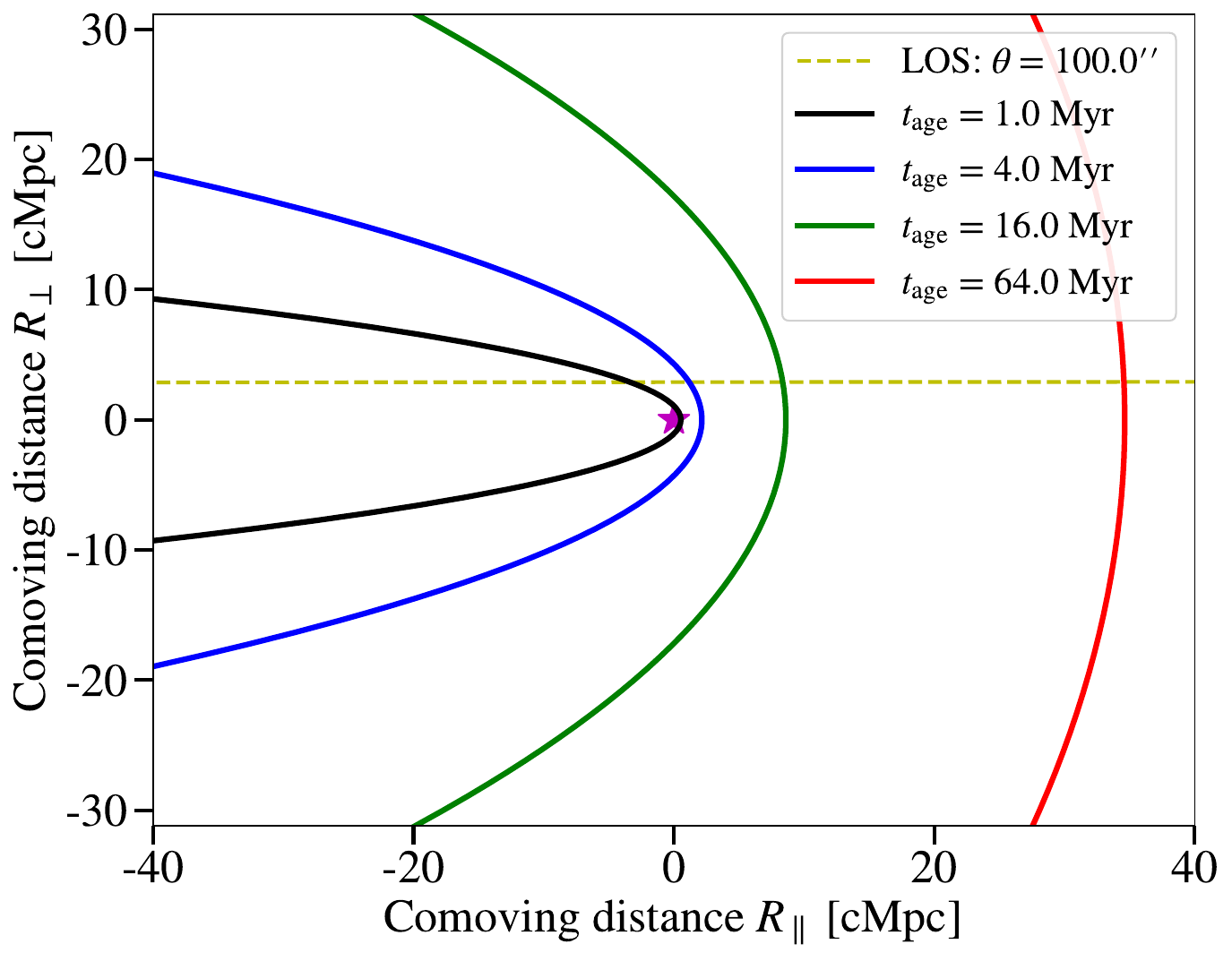}
  \end{center}
  \caption{\label{fig:LEs} Quasar light echo surfaces for various quasar ages ($z_{q}=2.5$). Each color corresponds to a different quasar age: $t_{\rm age} =$ 1.0 (black), 4.0 (blue), 16.0 (green), and 64.0 (red). The yellow dashed line is a line-of-sight direction with the angle of $\theta=100.0''$ .}
\end{figure}

\section{Light paths and quasar light echo} \label{sec:light_cone} 

To begin with, let us consider a quasar at a redshift of $z = z_{q}$ whose lifetime is $t=t_{q}$, as a working example. We assume that the quasar turned on at $t = 0$ and that we today observe photons that were emitted from the quasar at $t=t_{\rm age}(< t_{q})$. In addition to the photons directly propagated from the quasar ({\it direct}
photons), we should simultaneously observe Ly$\alpha$ photons that were  quasar photons emitted at $t<t_{\rm age}$ in the surrounding IGM ({\it induced} photons). We will consider light paths for the induced Ly$\alpha$ photons in the following sections. Note that since null geodesics are invariant under a conformal transformation \citep[e.g.,][]{Wald}, we use the conformal time, $\mathrm{d}\eta$, as well as the physical time $\mathrm{d}t\ (=a\mathrm{d}\eta)$.

\subsection{Light paths of quasar-induced Ly$\alpha$ photons \label{sec:light_path}}

Fig.~\ref{fig:schematic_LE} shows a schematic view where photons that were emitted by a quasar at $t=t_{\rm emi}(< t_{\rm age})$ induce Ly$\alpha$ photons, through resonant scattering or fluorescence, at the point of $R$ in the direction of $\theta$ with respect to the line of sight from an observer of $O$ to the quasar (blue solid line). $\chi_{q}$, $\chi_{r}$, and $\chi_{qr}$ are the comoving distances between the two points: $OQ$, $OR$, and $QR$, respectively. Since the direct photons ($Q \to O$) and the induced Ly$\alpha$ photons ($Q \to R \to O$) are observed at the same time, the relation between the comoving distance of light path and the comoving distance to the quasar is described as 
\begin{eqnarray}
	\Delta \eta_{\rm emi} + \chi_{qr} + \chi_{r} = \Delta \eta_{\rm age} + \chi_{q}, \label{eq:light_path}
\end{eqnarray}
where the time intervals $\Delta \eta_{\rm emi}$ and $\Delta \eta_{\rm age}$ correspond to the emission time $t_{\rm emi}$ and the quasar age $t_{\rm age}$ in the conformal time, respectively:
\begin{eqnarray}
	t_{\rm emi} = \int_{\eta_{\rm on}}^{\eta_{\rm emi}}a(\eta)\mathrm{d}\eta,  \quad
	t_{\rm age} = \int_{\eta_{\rm on}}^{\eta_{q}}a(\eta)\mathrm{d}\eta. \label{eq:t_def}
\end{eqnarray}
Here $\eta_{\rm on} \equiv \eta|_{t=0}$ is the conformal time at which the quasar turned on, $\eta_{\rm emi} \equiv \eta_{\rm on} + \Delta \eta_{\rm emi}$ is the conformal time at which Ly$\alpha$ photon were emitted from the quasar, and $\eta_{q} \equiv \eta_{\rm on} + \Delta \eta_{\rm age} = \eta_{0} - \chi_{q}$ is the conformal time corresponding to the comoving distance $\chi_{q}$ ($\eta_{0}$ is the conformal time of the present epoch). If the time interval $\Delta \eta_{\rm emi}$ is fixed in equation~(\ref{eq:light_path}), i.e., the sum of $\chi_{qr}$ and $\chi_{r}$ is constant, the point of $R$ is always located on an elliptic arc.

\subsection{Quasar light echo \label{sec:light_echo}}

The most interesting is the case of $\Delta \eta_{\rm emi} =0$, where the Ly$\alpha$ photons were induced by the photons leaving from the quasar at the same time, $t=0$, as the direct photons. In this case, the reflection point, $R_{\rm le}$, is located on the outermost elliptic arc, which we will hereafter call the {\it quasar light echo} surface\footnote{Although the term ``quasar light echo'' was originally introduced to expressly refer to the transverse proximity effect by \citet{2008ApJ...674..660V}, we here use it to express the radiation front rather than the ionization front.} and corresponds to the frontier of quasar radiation that we observe along with the direct photons (red solid line in Fig.~\ref{fig:schematic_LE}).

In the case of a spatially flat universe, substituting equation~(\ref{eq:light_path}) with $\Delta \eta_{\rm emi} =0$ into the law of cosines
\begin{eqnarray}
    \chi_{qr}^2 = \chi_{r}^2 + \chi_{q}^2 - 2\chi_{r} \chi_{q} \cos{\theta},   \label{eq:L_qr}
\end{eqnarray}
we obtain the following expression for the comoving distance to the light echo surface $\chi_{r,{\rm le}}$ as a function of $\theta$, $z_{q}$, and $t_{\rm age}$ (i.e., $\Delta \eta_{\rm age}$):
\begin{eqnarray}
    \chi_{r,{\rm le}}(\theta, z_{q}, t_{\rm age})  = \frac{\Delta \eta_{\rm age}}{\Delta \eta_{\rm age} + \chi_{q}(1-\cos{\theta})} \left(\chi_{q} + \frac{\Delta \eta_{\rm age}}{2}\right).  \label{eq:chi_le}
\end{eqnarray}
Here we define the redshift at the point of $R$ (or $R_{\rm le}$) as $z_{r}$ (or $z_{r,{\rm le}}$). Since the redshift $z_{r,{\rm le}}$ corresponds to the comoving distance $\chi_{r,{\rm le}}$, we can describe it as a function of $\theta$, $z_{q}$, and $t_{\rm age}$ in the same way as $\chi_{r,{\rm le}}$: $z_{r,{\rm le}}(\theta, z_{q}, t_{\rm age})$. Fig.~\ref{fig:LEs} shows the light echo surfaces for a quasar at
$z_{q}=2.5$, but assuming different ages. We find that, for a line of sight ($\theta = 100^{\prime \prime}$; dashed line) as an example, the intersection with the light echo surface moves to higher redshifts along the line of sight as the quasar age becomes older.

\section{Intergalactic medium, quasars, and star-forming galaxies} \label{sec:igm_qso}

Next, we introduce simple prescriptions for modeling the IGM, quasar radiation, and the clustering of star-forming galaxies. As we will see in Sections~\ref{sec:scattering} and \ref{sec:fluorescence}, the column density of HI gas clouds is a key ingredient to discuss the contribution of resonant scattering or fluorescence. We then adopt a typical column density distribution function in the IGM and consider its enhancement around quasars. As for quasar radiation, we assume a typical quasar spectrum and isotropic radiation. In addition, we will also discuss the attenuation due to the IGM in Section~\ref{sec:attenuation}. We also present a model based on the two-point correlation function of the large-scale dark matter distribution to account for the contribution of star-forming galaxies clustered around quasars to the Ly$\alpha$ emission excess.

\subsection{HI column density distribution \label{sec:f_N_HI}}

For a model describing the IGM, we rely on the distribution function of intergalactic absorbers introduced by \citet{2014MNRAS.442.1805I}, which is a function of redshift $z$ and the column density of neutral hydrogen $N_{\rm HI}$, composed of two components:
\begin{eqnarray}
    \frac{\partial^2 n}{\partial z \partial N_{\rm HI}}
        = f_{\rm LAF}(z)g_{\rm LAF}(N_{\rm HI}) + f_{\rm DLA}(z)g_{\rm DLA}(N_{\rm HI}),  
\end{eqnarray}
where 
\begin{eqnarray}
    g_{i}(N_{\rm HI})
        = B_{i}N_{\rm HI}^{-\beta_{i}}\mathrm{e}^{-N_{\rm HI}/N_{\rm c}},
        \label{eq:dn_dzdN}
\end{eqnarray}
the subscript $i = \{{\rm LAF, DLA}\}$, $\beta_{i}$ and $B_{i}$ are the power-law index and the normalization, respectively, for each component, and $N_{\rm c}$ is the cut-off column density for both. Note that each component was named after the Ly$\alpha$ forest (LAF) or damped Ly$\alpha$ systems (DLAs) reflecting its dominant contribution, however, both components contribute to Lyman limit systems (LLSs) corresponding to the middle of those two systems. In Appendix~\ref{app:f_N_HI}, we give the explicit expression for $f_{i}(z)$ and the model parameters assumed in the following.

In the following sections, we consider two extreme cases where HI gas clouds in the IGM are optically thin and thick to photons with different energies: Ly$\alpha$ and Lyman limit photons. We then define optically thick absorbers so that the corresponding optical depth is larger than $2$:\footnote{We follow \citet[][]{2013ApJ...776..136P} and set the optically depth criterion to 2, assuming we use their model parameters for the quasar-absorber cross-correlation function (see equation~\ref{eq:incidence_QSO}).}
\begin{align}
    \qquad
    N_{\rm HI} > N_{\rm Ly\alpha} &= 10^{13.5} {\rm cm^{-2}}\ (\text{optically thick to Ly$\alpha$})
    \nonumber \\ 
    N_{\rm HI} > N_{\rm LL} &= 10^{17.5} {\rm cm^{-2}}\
    (\text{optically thick to Lyman limit}),
    \nonumber \\ 
    \label{eq:def_thick_LAF} 
\end{align}
which are equivalent to $\sigma_{s}(\nu_{\alpha}) N_{\rm HI}>2$ and $\sigma_{\rm ion}(\nu_{\rm LL}) N_{\rm HI}>2$, respectively ($\sigma_{s}$ and $\sigma_{\rm ion}$ are the cross sections for Ly$\alpha$ resonant scattering and photoionization of HI atoms, respectively; see equations~(\ref{eq:sigma_s}) and (\ref{eq:sigma_pi}) for details). Based on the above definition, we introduce the incidence of optically thick absorbers as the number of them per line-of-sight physical length, $\mathrm{d}s=a\mathrm{d}\chi$,
\begin{align}
    \qquad
    l_{\rm IGM}^{\rm LAF}(z) &= \frac{\mathrm{d}z}{\mathrm{d}s}\int^{N_{\rm u}}_{N_{\rm Ly\alpha}} \frac{\partial^2 n}{\partial z \partial N_{\rm HI}} \mathrm{d}N_{\rm HI} \ (\text{for the LAF})
    \nonumber \\ 
    l_{\rm IGM}^{\rm LLS}(z) &= \frac{\mathrm{d}z}{\mathrm{d}s}\int^{N_{\rm u}}_{N_{\rm LL}} \frac{\partial^2 n}{\partial z \partial N_{\rm HI}} \mathrm{d}N_{\rm HI} \ (\text{for LLSs}),  \label{eq:incidence} 
\end{align}
where $\mathrm{d}z/\mathrm{d}s = (1+z)H(z)/c$ with the Hubble parameter $H$. We note that these incidences based on the column density distribution of equation~(\ref{eq:dn_dzdN}) reflect a randomly selected region in the Universe, which is emphasized by the subscript “IGM”. 

\begin{figure}
  \begin{center}
   \includegraphics[width=\linewidth]{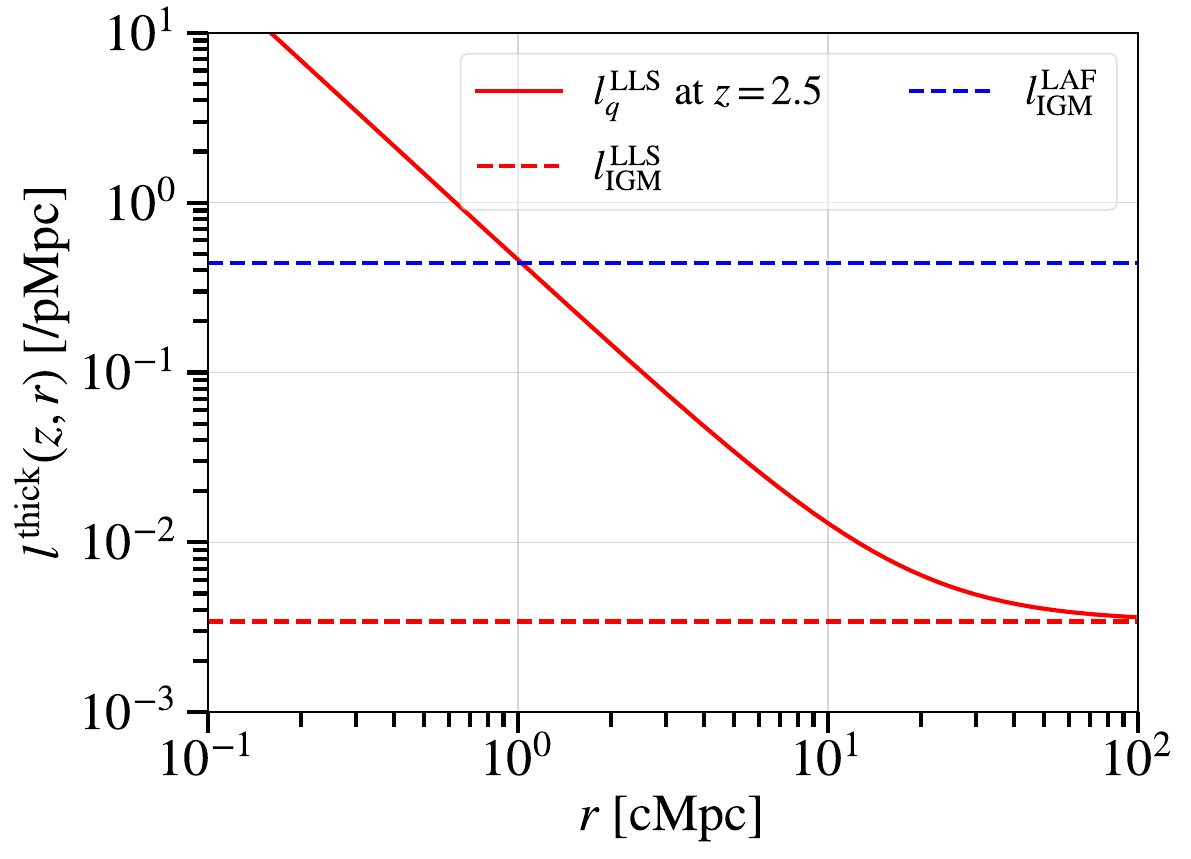}
  \end{center}
  \caption{
  \label{fig:incidence_thick} Incidence of optically thick objects per physical length at $z=2.5$. The dashed horizontal lines are the incidences for the LAF (blue) and LLSs (red), which are optically thick to Ly$\alpha$ and ionizing photons, respectively. We have also shown the incidence of LLSs around quasars, which is enhanced in the dense environments, in the red solid line. 
}
\end{figure}

However, in our context, we focus on the proximity of quasars and need to consider the effect of denser environments which are locally created by quasar halos. In particular, quasars have a relatively high linear bias factor \citep[e.g., $b_{q} = 3.64$ at a redshift of $z = 2.38$ as discussed in][]{2013JCAP...05..018F}, and therefore reside in high-density regions. Moreover, observations of absorption lines in background quasar spectra suggest that populations of optically thick HI absorbers are enhanced around the foreground quasar, even at a separation of $\sim 1 \ {\rm pMpc}$ \citep{2006ApJ...651...61H,2013ApJ...776..136P}. This means that although the fluorescent Ly$\alpha$ emission from each optically thick gas cloud does not depend on the HI number density within the cloud (see equation~\ref{eq:phi_UVB_ll}), the total contribution around quasars is further increased by a larger number of optically thick gas clouds in the dense environment \citep[e.g.,][]{2016ApJ...822...84M}. We then follow the prescription of \citet[][]{2013ApJ...776..136P} and take account of the enhancement in the incidence of absorbers around quasars by using the cross-correlation function between quasars and absorbers, $\xi_{\rm QA}$, with the separation (comoving) distance $r$:
\begin{eqnarray}
    l_{q}(z, r) = l_{\rm IGM}(z)(1+\xi_{\rm QA}(r)), \label{eq:incidence_QSO}  
\end{eqnarray}
where $\xi_{\rm QA}(r) = (r/r_{0})^{-\gamma}$ with the correlation (comoving) length $r_{0}$ and the power-law index $\gamma$. \citet[][]{2013ApJ...776..136P} measured the Ly$\alpha$ absorption around quasars using projected quasar pairs and found that $r_{0} = 12.5h^{-1}\ {\rm cMpc}$ and $\gamma = 1.68$ for LLSs\footnote{They used the criterion $N_{\rm HI} > 10^{17.3} {\rm cm^{-2}}$ when defining LLSs, which means that our criterion, equation~(\ref{eq:def_thick_LAF}), is slightly more conservative.} by fitting it to the model. Hereafter, we use these values to estimate the cross-correlation for LLSs. Fig~\ref{fig:incidence_thick} shows the incidence of LLSs around quasars, equation~(\ref{eq:incidence_QSO}), as a function of the distance $r$ (red solid line). We can clearly see the enhancement in the inner region from the incidence in the average IGM (red dashed line). We would like to note that the LAF or optically thin absorbers are also enhanced around quasars. However, we consider the clustering enhancement only for LLSs because of their dominant contribution to the Ly$\alpha$ emission excess (see Section~\ref{sec:results} for details) and the lack of measurements for the incidence of low column density absorbers on small scales.

\begin{figure*}
  \begin{center}
   \includegraphics[width=0.90\linewidth]{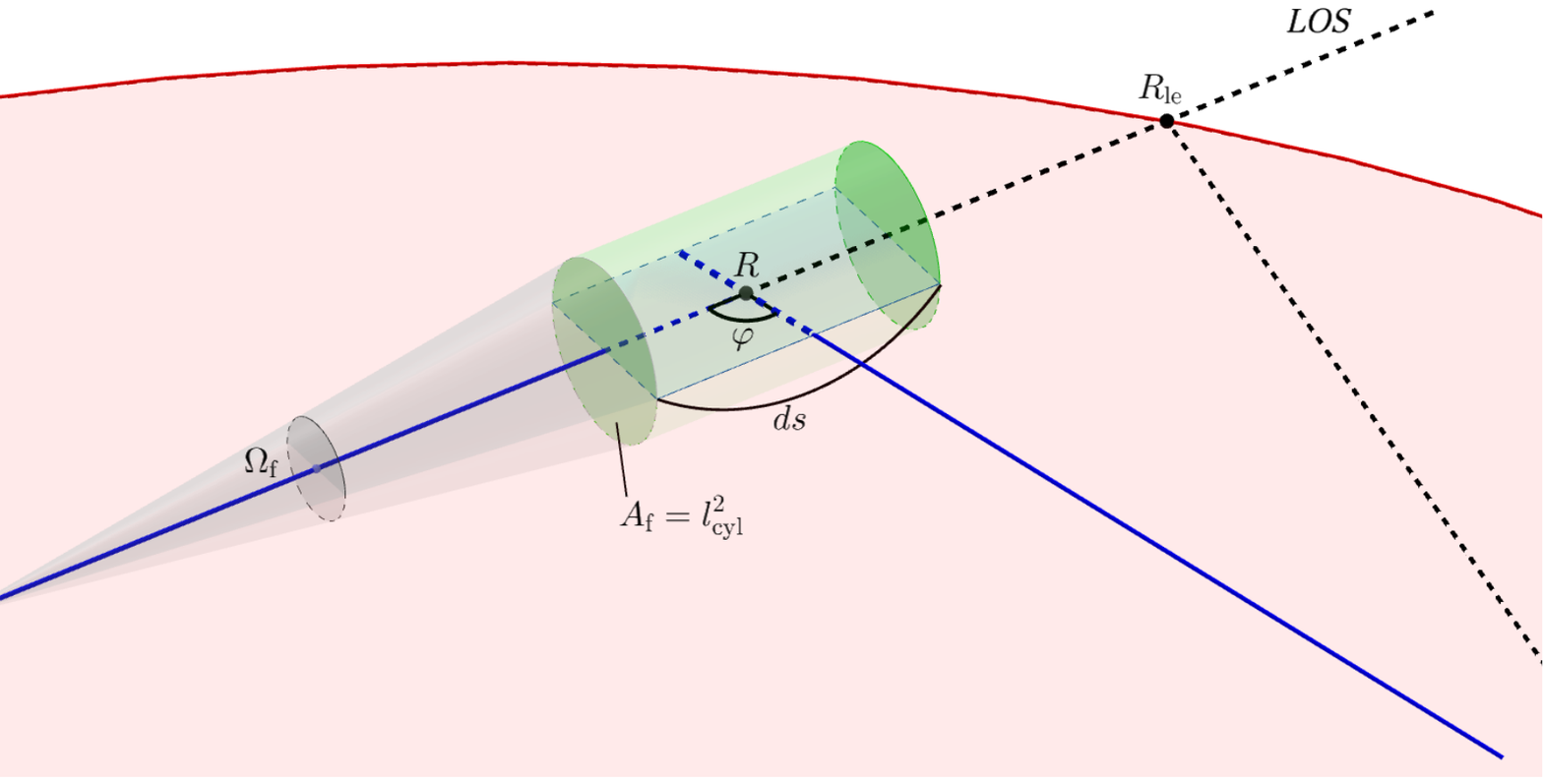}
  \end{center}
  \caption{\label{fig:around_R} Quasar-induced Ly$\alpha$ emission observed by a spectroscopic fiber with an aperture of $\Omega_{\rm f}$. Each line corresponds to those in Fig.~\ref{fig:schematic_LE}.}
\end{figure*}

\subsection{Quasar spectrum}

To estimate a typical flux of quasars in intergalactic space, we adopt the average spectrum presented by \citet{2015MNRAS.449.4204L}, who analyzed 53 quasars at high redshift ($z \simeq 2.4$) from the Hubble Space Telescope survey for Lyman limit absorption systems and obtained the following spectral index of the continuum slope:  
\begin{eqnarray}
    \alpha = 
\begin{cases}
    -0.61 \quad (\nu < \nu_{\rm LL})\\
    -1.70 \quad (\nu > \nu_{\rm LL}).
\end{cases}
\end{eqnarray}
Using this spectral index, the quasar luminosity density $L_{q}(\nu_{q})$ (in $\rm{erg}\ \rm{s}^{-1}\ \rm{Hz}^{-1}$) is described by
\begin{eqnarray}
    L_{q}(\nu_{q}) = L_{q,\nu_{\rm LL}} \left(\frac{\nu_{q}}{\nu_{\rm LL}}\right)^{\alpha} 
\end{eqnarray}
where $\nu_{q}$ is the rest-flame frequency at the quasar position and $L_{q,\nu_{\rm LL}}$ is the luminosity density at the Lyman limit. We set the fiducial value of $L_{q,\nu_{\rm LL}}$ to $1.0 \times 10^{31}\ \rm{erg}\ \rm{s}^{-1}\ \rm{Hz}^{-1}$ \citep[e.g.,][]{2010ApJ...708.1048K}, which corresponds to an absolute magnitude at rest-frame 1450\AA: $M_{1450}\simeq -26$.

Then, the quasar flux per unit frequency at the point of $R$ in Fig.~\ref{fig:around_R}, denoted as $F_{q}(\nu)$ (in $\rm{erg}\ \rm{cm}^{-2}\ \rm{s}^{-1}\ \rm{Hz}^{-1}$), is
\begin{eqnarray}
    F_{q}(\nu) = (1+z_{\rm rel}) \frac{L_{q}\left((1+z_{\rm rel})\nu \right)}{4\pi d_{L,\rm rel}^2}.
\end{eqnarray}
Here $z_{\rm rel}$ is the relative redshift of the quasar with respect to the point of $R$ and defined by $1 + z_{\rm rel} \equiv a(\eta_{r})/a(\eta_{\rm emi})$, where $\eta_{r}$ is the conformal time corresponding to the comoving distance $\chi_{r}$: $\eta_{r} = \eta_{0} - \chi_{r}$. $d_{L,\rm rel}$ is the relative luminosity distance defined in the same way. Since $\eta_{\rm emi} (= \eta_{r} - \chi_{qr})$ is a function of $\theta$, $z_{r}$, and $z_{q}$ (see equation~\ref{eq:L_qr}), the quasar flux $F_{q}$ implicitly depends on these parameters: $F_{q}(\nu; \theta, z_{r}, z_{q})$. Note that the additional factor $(1+z_{\rm rel})$, by which the conventional expression of flux is multiplied, is introduced because $L_{q}(\nu)$ (or $F_{q}(\nu)$) is the luminosity (or flux) {\it density} rather than the total luminosity (or flux).

\subsection{Clustering of star-forming galaxies \label{sec:SFG}}

While in this paper, we focus on assessing the effect of quasar radiation on the Ly$\alpha$ emission from the surrounding IGM, the clustering enhancement of star-forming galaxies in the dense environment around quasars also boosts the Ly$\alpha$ emission surface brightness. A study of the environment of quasars at $z=3{\rm-}4.5$ from the MUSE survey found more LAEs around quasars, which are presumably star-forming galaxies, and suggested that the excess is mainly caused by the local denser environment and partly by the quasar radiation \citep[][]{2021MNRAS.503.3044F}. Moreover, \citet{2022ApJS..262...38L} recently performed the Ly$\alpha$ emission intensity mapping with LRG spectra from the eBOSS data and compared it separately with two different models, where Ly$\alpha$ sources are star-forming galaxies or quasars that trace the large-scale structure. The results suggest that the star-forming galaxy-based model is preferred to interpret the observed quasar-Ly$\alpha$ emission cross-correlation on scales of $\gsim 1\ {\rm cMpc}$, since the quasar-based model requires an extremely high Ly$\alpha$ luminosity per quasar, $\gsim 10^{45}\ {\rm erg}\ {\rm s}^{-1}$, which is comparable with or even brighter than previously observed ultraluminous quasars. Therefore, we need to consider quasars and their surrounding external radiation sources simultaneously to understand their proportions or scale dependencies. We then adopt a methodology based on linearly biased dark matter density fluctuations to model the contribution of star-forming galaxies, following \citet{2016MNRAS.457.3541C} and \citet{2022ApJS..262...38L}. Assuming the large-scale fluctuations of Ly$\alpha$ emission intensity is due to the large-scale clustering of star-forming galaxies (i.e., $\mu_{\alpha} \propto \rho^{\rm SFG}_{\alpha}$, where $\mu_{\alpha}$ is the Ly$\alpha$ emission surface brightness and $\rho^{\rm SFG}_{\alpha}$ is the comoving Ly$\alpha$ luminosity density of star-forming galaxies), the cross-correlation between quasars and Ly$\alpha$ emissions in the linear regime \citep{1992ApJ...385L...5H} is 
\begin{eqnarray}
    \xi_{{\rm q}\alpha}(s, \mu) = b_{q}b_{\alpha}\langle \mu_{\alpha} \rangle \xi(s, \mu),
\end{eqnarray}
where $s$ is the redshift-space (comoving) separation between quasars and pixels including Ly$\alpha$ emissions and $\mu = s_{\parallel}/s$ with its line-of-sight component $s_{\parallel}$. Here $b_{q}$ and $b_{\alpha}$ are the linear bias factors for quasars and Ly$\alpha$ emissions, respectively, $\langle \mu_{\alpha} \rangle$ is the mean Ly$\alpha$ emission surface brightness, and the redshift-space correlation function of the dark matter distribution, $\xi(s, \mu)$, can be expanded using the Legendre polynomials $\mathcal{L}_{\ell}(\mu)$, 
\begin{eqnarray}
    \xi(s, \mu) = \sum_{\ell=0,2,4}\xi_{\ell}(s)\mathcal{L}_{\ell}(\mu),
\end{eqnarray}
with
\begin{eqnarray}
    &&\xi_{0}(s) = \left(1+\frac{1}{3}(\beta_{q}+\beta_{\alpha})+\frac{1}{5}\beta_{q}\beta_{\alpha}\right)\xi(r),
    \nonumber \\
    &&\xi_{2}(s) = \left(\frac{2}{3}(\beta_{q}+\beta_{\alpha})+\frac{4}{7}\beta_{q}\beta_{\alpha}\right)\left[\xi(r)-\bar{\xi}(r)\right],
    \nonumber \\
    &&\xi_{4}(s) = \frac{8}{35}\beta_{q}\beta_{\alpha}\left[\xi(r)+\frac{5}{2}\bar{\xi}(r)-\frac{7}{2}\dbar{\xi}(r)\right],
    \label{eq:multipoles}
\end{eqnarray}
where 
\begin{eqnarray}
    &&\bar{\xi}(r) = \frac{3}{r^{3}}\int^{r}_{0}\xi(r')r'^{2}\mathrm{d}r',
    \nonumber \\
    &&\dbar{\xi}(r) = \frac{5}{r^{5}}\int^{r}_{0}\xi(r')r'^{4}\mathrm{d}r',
\end{eqnarray}
and $\beta_{x} = f/b_{x}$ with the linear growth rate $f$. We then compute the {\it truncated} multiples to cut off nonlinear effects on small scales, such as random pairwise motions \citep[e.g.,][]{2014MNRAS.444..476R,2016MNRAS.458.1948M,2019MNRAS.487.2424M},\footnote{Note that if we can ignore the effect of random pairwise motions, the redshift-space separation $s$ is equal to the separation $r$ that appears in equations~(\ref{eq:incidence_QSO}) or (\ref{eq:multipoles}).}
\begin{eqnarray}
    \hat{\xi}_{\ell}(s) = \frac{2\ell+1}{2} \int^{\mu_{\rm max}}_{-\mu_{\rm max}} \xi(s, \mu) \mathcal{L}_{\ell}(\mu) \mathrm{d}\mu,
\end{eqnarray}
where $\mu_{\rm max} = \sqrt{1-(s_{\perp, {\rm min}}/s)^2}$, corresponding to limiting transverse separations to $s_{\perp} > s_{\perp, {\rm min}}$. In the following, we only focus on the isotropically averaged cross-correlation to compare with the contribution due to quasar radiation. We then define the contribution of star-forming galaxies to the Ly$\alpha$ emission excess as 
\begin{eqnarray}
    \xi^{\rm SFG}_{{\rm q}\alpha}(s)= b_{q}b_{\alpha}\langle \mu_{\alpha} \rangle \hat{\xi}_{0}(s) \label{eq:xi_SFG}
\end{eqnarray}
and follow \citet{2022ApJS..262...38L} to set the parameters: $s_{\perp, {\rm min}}=4h^{-1}\ {\rm cMpc}$, $b_{q} = 3.64$ \citep{2013JCAP...05..018F}, and $b_{\alpha} = 3$ \citep{2016MNRAS.457.3541C}. In addition, we evaluate the linear growth rate, $f$, with a good approximation for $\Lambda$CDM: $f\simeq\Omega_{m}(z)^{0.55}$ \citep{1998ApJ...508..483W,2005PhRvD..72d3529L}. We will discuss the adopted value for $\langle \mu_{\alpha} \rangle$ in Section~\ref{sec:forecast}.

\section{Resonant scattering of Lyman-alpha photons} \label{sec:scattering}
 
In this section, we consider how Ly$\alpha$ photons that were emitted by a quasar are resonantly scattered by HI atoms in the IGM around the quasar, and then estimate the observed flux. Here, we focus on the scattering point, i.e., the point of $R$ in Fig.~\ref{fig:schematic_LE}, and then $\lambda$ and $\nu$ represent the rest-frame wavelength and frequency at the point of $R$, respectively, unless noted otherwise. Accordingly, for instance, the term ``quasar Ly$\alpha$ photons'' refers to photons with the Ly$\alpha$ wavelength in the rest-frame of the point of $R$, which corresponds to continuum photons slightly blueward of Ly$\alpha$ in the quasar rest frame.

\subsection{Ly$\alpha$ volume emissivity \label{sec:j_Lya}}

Following the formalism that is presented by \citet{2013ApJ...766...58H}, we consider gas clouds with the number density of $n_{\rm c}$ and the cross-sectional area of $\sigma_{\rm c}$. From the quasar flux and the HI column density for each cloud $N_{\rm HI, c}$, we can define the reaction rate for the scattering of quasar Ly$\alpha$ photons for each cloud, per unit area (in $\rm{cm}^{-2}\ \rm{s}^{-1}$):
\begin{eqnarray}
     \Upsilon_{{\rm Q}(s)}(N_{\rm HI, c}) = \int \mathrm{d}\nu \frac{F_{q}(\nu)}{h\nu} (1-\mathrm{e}^{-\tau_{s, {\rm c}}}),  \label{eq:Ups_s}
\end{eqnarray}
where $h$ is the Planck constant and $\tau_{s, {\rm c}} = \sigma_{s} N_{\rm HI, c}$ is the optical depth per cloud. The cross section for Ly$\alpha$ scattering, $\sigma_{s}$, is described as a convolution of the Ly$\alpha$ scattering cross section for a single HI atom and the velocity distribution of the atoms \citep{Radipro, Mesinger}:
\begin{eqnarray}
     \sigma_{s}(\nu, T) = \frac{3\lambda_{\alpha}^{2}a_{\varv}}{2\sqrt{\pi}}H(a_{v}, x),  \label{eq:sigma_s}
\end{eqnarray}
with
\begin{eqnarray}     
     a_{\varv} \equiv \frac{A_{\alpha}}{4\pi\Delta \nu_{\alpha}}, \quad x \equiv \frac{\nu-\nu_{\alpha}}{\Delta \nu_{\alpha}},
     \nonumber
\end{eqnarray}
where $\lambda_{\alpha}$ and $\nu_{\alpha}$ are the Ly$\alpha$ wavelength and frequency, respectively, and $A_{\alpha}$ is the Einstein-A coefficient for the Ly$\alpha$ transition, and $\Delta \nu_{\alpha} \equiv \nu_{\alpha}\varv_{\rm th}/c = \nu_{\alpha}\sqrt{2k_{B}T/m_{p}}/c$ with the Boltzmann constant $k_{B}$. Here $H(a_{v}, x)$ is the Voigt function, which is normalized to $\sqrt{\pi}$, i.e. $\int H(a_{v}, x)\mathrm{d}x = \sqrt{\pi}$ (and $H(a_{v}, 0) = 1$).  

By assuming that quasar Ly$\alpha$ photons are isotopically scattered by HI atoms, now we can write down the Ly$\alpha$ volume emissivity (in $\rm{erg}\ \rm{s}^{-1}\ \rm{cm}^{-3}\ \rm{arcsec}^{-2}$):
\begin{eqnarray}     
     j_{\rm Ly\alpha}(N_{\rm HI, c}) = \frac{h\nu_{\alpha}}{4\pi}n_{\rm c}\sigma_{\rm c}\Upsilon_{{\rm Q}(s)}.  \label{eq:j_lya_s}
\end{eqnarray}
Let us consider two extreme cases: optically thin and thick limits. At the optically thin limit ($N_{\rm HI, c} \ll 1/\sigma_{s}$), 
\begin{eqnarray}     
     j_{\rm Ly\alpha} 
     &\to& \frac{h\nu_{\alpha}}{4\pi}n_{\rm c}\sigma_{\rm c}\int \mathrm{d}\nu \frac{F_{q}(\nu)}{h\nu} \tau_{s, {\rm c}}
     \nonumber \\
     &=& \frac{h\nu_{\alpha}}{4\pi}n_{\rm c}\sigma_{\rm c}\ N_{\rm HI, c}\int \mathrm{d}\nu \frac{F_{q}(\nu)}{h\nu} \sigma_{s}(\nu)
     \nonumber \\
     &=& \frac{h\nu_{\alpha}}{4\pi}\left(\frac{\mathrm{d}f_{\rm C}}{\mathrm{d}s} N_{\rm HI, c}\right)\ \Gamma_{q,s},   \label{eq:j_lya_s_thin}
\end{eqnarray}
where
\begin{eqnarray}     
     \Gamma_{q,s} &=& \int \mathrm{d}\nu \frac{F_{q}(\nu)}{h\nu} \sigma_{s}(\nu) 
     \nonumber \\
     &\simeq& \frac{F_{q}(\nu_{\alpha})}{h\nu_{\alpha}}(\sqrt{\pi}\Delta \nu_{\alpha})\sigma_{s}(\nu_{\alpha}) \label{eq:gamma_q_s}
\end{eqnarray}
(in $\rm{s}^{-1}$) is the scattering rate of quasar Ly$\alpha$ photons and $\mathrm{d}f_{\rm C}/\mathrm{d}s(=n_{\rm c}\sigma_{\rm c})$ is the covering factor per unit physical length. In the second line of equation~(\ref{eq:gamma_q_s}), we assumed that the spectrum $F_{q}(\nu)$ does not change rapidly around $\nu = \nu_{\alpha}$ since we are here only considering the continuum component for the quasar spectrum. We can see that the emissivity is described by the product of the number of HI atoms per volume and the scattering rate per HI atom.

On the other hand, in the optically thick case ($N_{\rm HI, c} \gg 1/\sigma_{s}$), equations~(\ref{eq:j_lya_s}) becomes 
\begin{eqnarray}     
     j_{\rm Ly\alpha} 
     &\to& \frac{h\nu_{\alpha}}{4\pi}n_{\rm c}\sigma_{\rm c}\ \frac{F_{q}(\nu_{\alpha})}{h} \int \frac{\mathrm{d}\nu}{\nu_{\alpha}} (1-\mathrm{e}^{-\tau_{s, {\rm c}}})
     \nonumber \\
     &=& \frac{h\nu_{\alpha}}{4\pi}\frac{\mathrm{d}f_{\rm C}}{\mathrm{d}s}\ \Phi_{q,{\rm Ly\alpha}}, \label{eq:j_lya_s_thick}
\end{eqnarray}
where
\begin{eqnarray}     
     \Phi_{q,{\rm Ly\alpha}} = \frac{F_{q}(\nu_{\alpha})}{h} \underbrace{\int \frac{\mathrm{d}\nu}{\nu_{\alpha}} (1-\mathrm{e}^{-\tau_{s, {\rm c}}})}_{=\ W\ (\text{dimensionless EW})} \label{eq:phi_q_lya}
\end{eqnarray}
(in $\rm{s}^{-1}\ \rm{cm}^{-2}$) is the number flux of quasar Ly$\alpha$ photons. The integral in equation~(\ref{eq:phi_q_lya}) corresponds to the dimensionless equivalent width (EW), $W$, which can be approximated as
\begin{eqnarray}     
     W \simeq \frac{2\varv_{\rm th}}{c}\sqrt{\ln \left( \frac{\tau_{s, {\rm c}}(\nu_{\alpha})}{\ln2} \right)}, \label{eq:W}
\end{eqnarray}
with $5\%$ accuracy for $1.254<\tau_{s, {\rm c}}(\nu_{\alpha})\lsim 10^{4}$ \citep[e.g.,][]{Draine}.

In order to compute the Ly$\alpha$ emissivities in both optically thin and thick cases, we need to evaluate $(\mathrm{d}f_{\rm C}/\mathrm{d}s) N_{\rm HI, c}$ and $\mathrm{d}f_{\rm C}/\mathrm{d}s$, respectively. While we assumed a specific value for $n_{\rm c}$ or $\sigma_{\rm c}$ in the above expressions, we should observe the sum of Ly$\alpha$ emissions due to gas clouds with different column densities. Recalling that the incidence of absorbers was defined as the sum of the contributions over a range of the HI column density (equation~\ref{eq:incidence}), we can then evaluate those two effects as follows: 
\begin{align}
    \qquad \qquad
    \sum_{\rm thick} \frac{\mathrm{d}f_{\rm C}}{\mathrm{d}s} &= l_{\rm IGM}^{\rm LAF}(z)
    \nonumber \\
    \sum_{\rm thin} \frac{\mathrm{d}f_{\rm C}}{\mathrm{d}s} N_{\rm HI, c} &= \frac{\mathrm{d}z}{\mathrm{d}s}\int^{N_{\rm Ly\alpha}}_{N_{\rm l}} \frac{\partial^2 n}{\partial z \partial N_{\rm HI}} N_{\rm HI} \mathrm{d}N_{\rm HI},
    \label{eq:def_thin_LAF} 
\end{align}
where the capital sigma symbol represents the summation of contributions over all optically thick(thin) gas clouds. For the integral of the second line, we note that high column density gas clouds ($\simeq N_{\rm Ly\alpha}$) are no longer optically thin, which leads to an overestimate of $(\mathrm{d}f_{\rm C}/\mathrm{d}s) N_{\rm HI, c}$. 

Furthermore, we need a gas temperature to calculate the scattering cross section (equation~\ref{eq:sigma_s}). 
The IGM temperature has been measured by focusing on the width of Ly$\alpha$ absorption lines and found to be $T \sim 10^{4}$K at the mean density in the redshift range of $2\lsim z\lsim 4$ \citep[e.g.,][]{2016ARA&A..54..313M}. Since the summations in the above equations mostly come from absorbers with $N_{\rm HI}\lsim N_{\rm Ly\alpha}$, we then adopt $T=10^{4}$K to estimate the velocity dispersion of gas, $\varv_{\rm th}$, in the following \citep[e.g.,][]{2001ApJ...559..507S}. The dimensionless EW of equation~(\ref{eq:W}) also depends on the HI column density through the optical depth. However, given the dominance of the lowest column density and the logarithmic dependence, we use the value of $(W\cdot\lambda_{\alpha}) = 0.11$\AA\ evaluated with the column density of $N_{\rm Ly\alpha}$.

\section{Lyman-alpha fluorescence} \label{sec:fluorescence}

As a contribution to Ly$\alpha$ emission excess around quasars, we have discussed the resonant scattering of quasar Ly$\alpha$ photons by HI atoms in the surrounding gas clouds. Here we move on to estimate the effect of Ly$\alpha$ fluorescence in the same gas clouds, which is induced by ionizing photons from local quasars. Before that, we start with the Ly$\alpha$ fluorescence due to the UV background for a more general case.

\subsection{Fluorescence due to the UV background}

Considering the UV background of the intensity $J_{\rm UV}(\nu)$ (in $\rm{erg}\ \rm{cm}^{-2}\ \rm{s}^{-1} \rm{sr}^{-1} \rm{Hz}^{-1}$), the reaction rate for the ionization of HI atoms in each cloud, per unit area (see equation~\ref{eq:Ups_s}):
\begin{eqnarray}
     \Upsilon_{{\rm UVB}({\rm ion})} = \int_{\nu_{\rm LL}}^{\infty} \mathrm{d}\nu \frac{4 \pi J_{\rm UV}(\nu)}{h\nu} (1-\mathrm{e}^{-\tau_{{\rm ion}, {\rm c}}}),  \label{eq:Ups_i}
\end{eqnarray}
where $\tau_{{\rm ion}, {\rm c}} = \sigma_{\rm ion} N_{\rm HI, c}$ is the optical depth per cloud and $\sigma_{\rm ion}(\nu)$ is the cross section for photoionization of HI atoms, which is defined by
\begin{eqnarray}
    \sigma_{\rm ion} = \sigma_{\nu_{\rm LL}} (\nu/\nu_{\rm LL})^{-3}.  \label{eq:sigma_pi}
\end{eqnarray}
Here $\nu_{\rm LL}$ is the frequency at the Lyman limit and $\sigma_{\nu_{\rm LL}}$ is the cross section at the threshold, which is given by $\sigma_{\nu_{\rm LL}} = 6.3\times 10^{-18}\ \rm{cm}^2$ \citep{Osterbrock, Draine}. While in photoionization equilibrium, ionization and recombination rates are equal to each other, each recombination event does not necessarily produce a single Ly$\alpha$ photon because there are some types of radiative cascades not resulting in Ly$\alpha$. For regions that are optically thin (thick) to ionizing radiation, with a temperature around $10^{4}$K, a fraction $\eta_{\rm thin} \simeq 41.0\%$ ($\eta_{\rm thick} \simeq 68.6\%$) of recombinations emit a Ly$\alpha$ photon \citep[e.g.,][]{1996ApJ...468..462G,2013ApJ...766...58H,2014PASA...31...40D}. Then, by analogy to the resonant scattering case (equation~\ref{eq:j_lya_s}), the Ly$\alpha$ volume emissivity is
\begin{eqnarray}     
    j_{\rm Ly\alpha} &=& \frac{h\nu_{\alpha}}{4\pi}n_{\rm c}\sigma_{\rm c}\ \eta_{\rm X}\Upsilon_{{\rm UVB}({\rm ion})},  \label{eq:j_lya_f}
\end{eqnarray}
where $\eta_{\rm X} = \{\eta_{\rm thin},\ \eta_{\rm thick}\}$.

Recalling equation~(\ref{eq:j_lya_s_thin}), for the the optically thin limit ($N_{\rm HI, c} \ll 1/\sigma_{\rm ion}$), the above equation becomes 
\begin{eqnarray}     
     j_{\rm Ly\alpha} 
     &\to& \frac{h\nu_{\alpha}}{4\pi}n_{\rm c}\sigma_{\rm c}\ \eta_{\rm thin} N_{\rm HI, c}\int_{\nu_{\rm LL}}^{\infty} \mathrm{d}\nu \frac{4 \pi J_{\rm UV}(\nu)}{h\nu} \sigma_{\rm ion}(\nu)
     \nonumber \\
     &=& \frac{h\nu_{\alpha}}{4\pi}\eta_{\rm thin}\ \left(\frac{\mathrm{d}f_{\rm C}}{\mathrm{d}s} N_{\rm HI, c}\right)\ \Gamma_{{\rm UVB},{\rm ion}},  \label{eq:j_lya_f_thin}
\end{eqnarray}  
where
\begin{eqnarray}     
     \Gamma_{{\rm UVB},{\rm ion}} &=& \int_{\nu_{\rm LL}}^{\infty} \mathrm{d}\nu \frac{4 \pi J_{\rm UV}(\nu)}{h\nu} \sigma_{\rm ion}(\nu)
     \label{eq:gamma_UVB_i}
\end{eqnarray}
(in $\rm{s}^{-1}$) is the ionization rate due to the UV background. In the optically thick case ($N_{\rm HI, c} \gg 1/\sigma_{\rm ion}$), from equations~(\ref{eq:j_lya_s_thick}), we obtain the following expression: 
\begin{eqnarray}     
     j_{\rm Ly\alpha} 
     &\to& \frac{h\nu_{\alpha}}{4\pi}n_{\rm c}\sigma_{\rm c}\ \eta_{\rm thick} \int_{\nu_{\rm LL}}^{\infty} \mathrm{d}\nu\frac{4 \pi J_{\rm UV}(\nu)}{h\nu}
     \nonumber \\
     &=& \frac{h\nu_{\alpha}}{4\pi}\eta_{\rm thick}\ \frac{\mathrm{d}f_{\rm C}}{\mathrm{d}s}\ \Phi_{{\rm UVB},{\rm LL}}, \label{eq:j_lya_f_thick}
\end{eqnarray}
where
\begin{eqnarray}     
     \Phi_{{\rm UVB},{\rm LL}} = \int_{\nu_{\rm LL}}^{\infty} \mathrm{d}\nu\frac{4 \pi J_{\rm UV}(\nu)}{h\nu} \label{eq:phi_UVB_ll}
\end{eqnarray}
(in $\rm{s}^{-1}\ \rm{cm}^{-2}$) is the number flux of Lyman limit photons of the UV background. 

In the following, we use the UVB ionization rate of $\Gamma_{{\rm UVB},{\rm ion}} = 9 \times 10^{-13}\ \rm{s}^{-1}$ in the redshift range of $2\lsim z\lsim 3$ \citep[][]{2012ApJ...746..125H}. In addition, we derive $\Phi_{{\rm UVB},{\rm LL}}$ from the same $\Gamma_{{\rm UVB},{\rm ion}}$ using the gray absorption cross-section of $2.6 \times 10^{-18}\ \rm{cm}^{2}$, which is defined by the ratio of $\Gamma_{{\rm UVB},{\rm ion}}$ to $\Phi_{{\rm UVB},{\rm LL}}$ \citep[][]{2013MNRAS.430.2427R}. Also in this fluorescence case, we can evaluate $(\mathrm{d}f_{\rm C}/\mathrm{d}s) N_{\rm HI, c}$ and $\mathrm{d}f_{\rm C}/\mathrm{d}s$ in the same way as equation~(\ref{eq:def_thin_LAF}):
\begin{align}
    \qquad \qquad
    \sum_{\rm thick} \frac{\mathrm{d}f_{\rm C}}{\mathrm{d}s} &= l_{\rm IGM}^{\rm LLS}(z) \ \left({\rm or}\ l_{q}^{\rm LLS}(z, r)\right)
    \nonumber \\
    \sum_{\rm thin} \frac{\mathrm{d}f_{\rm C}}{\mathrm{d}s} N_{\rm HI, c} &= \frac{\mathrm{d}z}{\mathrm{d}s}\int^{N_{\rm LL}}_{N_{\rm l}} \frac{\partial^2 n}{\partial z \partial N_{\rm HI}} N_{\rm HI} \mathrm{d}N_{\rm HI}.
    \label{eq:def_thin_LLS} 
\end{align}

\subsection{Fluorescence in the vicinity of a local quasar \label{sec:F_Q_f}}

The fluorescent Ly$\alpha$ emissivity in equation~(\ref{eq:j_lya_f}) is defined globally (i.e., not considering the photoionization of a local quasar) and then should be regarded as the ``fluorescence'' background that is induced by the UV background. Therefore, when we consider the Ly$\alpha$ fluorescence in the vicinity of a local quasar, we need to add the contribution of ionizing photons emitted from the quasar, which is given by replacing $4 \pi J_{\rm UV}(\nu)$ with $F_{q}(\nu)$ in equation~(\ref{eq:Ups_i}). In this case, the Ly$\alpha$ emissivities in the optically thin and thick cases (analogy to equations~\ref{eq:j_lya_f_thin} and \ref{eq:j_lya_f_thick}) are described by the ionization rate and the number flux of Lyman limit photons due to the local quasar, respectively:
\begin{eqnarray}     
     \Gamma_{q,{\rm ion}} &=& \int_{\nu_{\rm LL}}^{\infty} \mathrm{d}\nu \frac{F_{q}(\nu)}{h\nu} \sigma_{\rm ion}(\nu),
     \label{eq:gamma_q_i}
     \\
     \Phi_{q,{\rm LL}} &=& \int_{\nu_{\rm LL}}^{\infty} \mathrm{d}\nu\frac{F_{q}(\nu)}{h\nu}. \label{eq:phi_q_ll}
\end{eqnarray}
We note that in this and the previous sections, we have only considered quasars as radiation sources of Ly$\alpha$ emission from the IGM. As we discussed in Section~\ref{sec:SFG}, Ly$\alpha$ emissions from clustered star-forming galaxies directly contribute to their excess around quasars. In addition, Ly$\alpha$ and ionizing photons from star-forming galaxies can be radiation sources of the Ly$\alpha$ emission from the IGM through resonant scattering and fluorescence, respectively (see also Section~\ref{sec:potential_contributions}).

\begin{figure}
  \begin{center}
   \includegraphics[width=\linewidth]{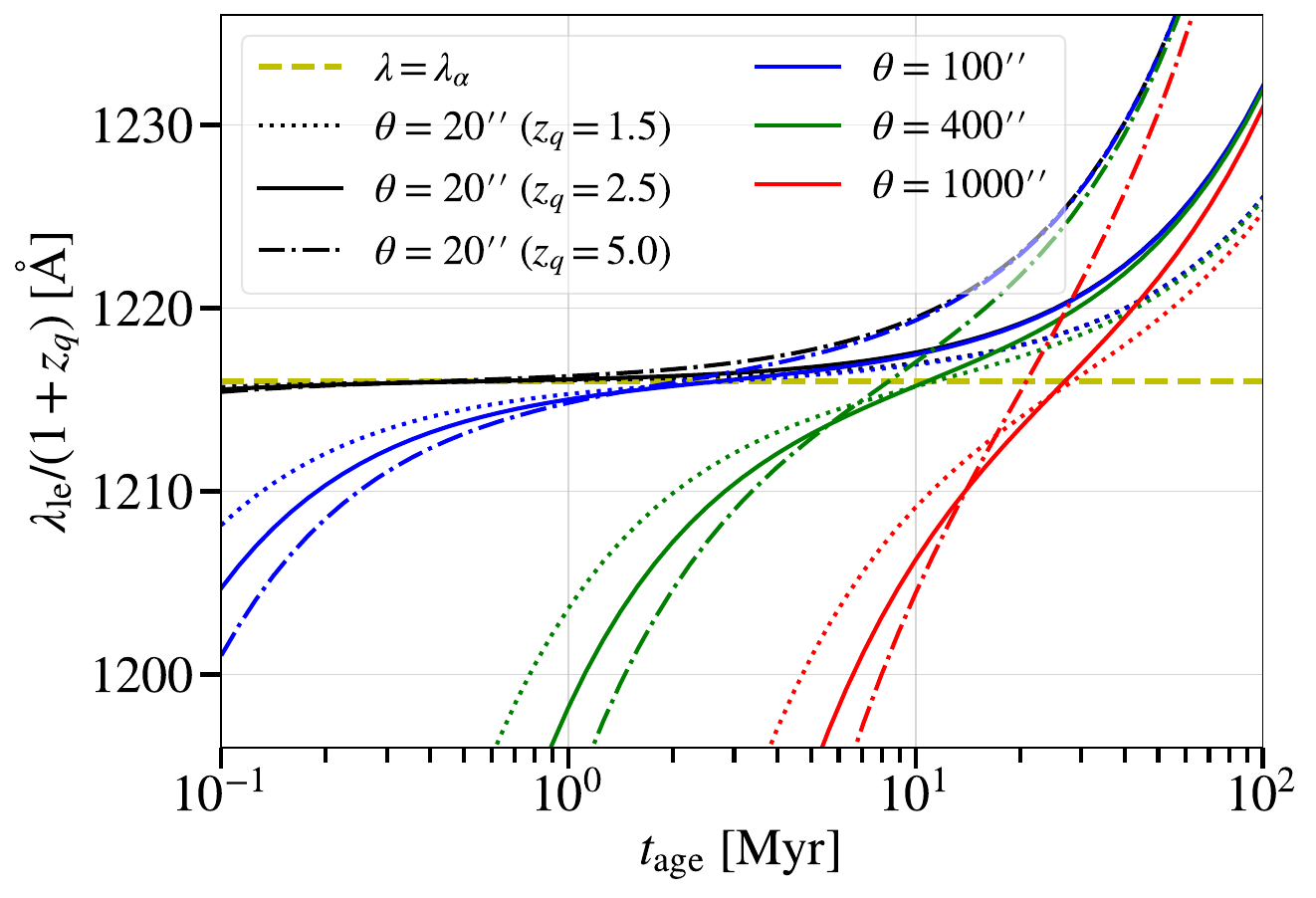}
  \end{center}
  \caption{\label{fig:lambda_le} Upper limit of the scattered Ly$\alpha$ wavelength in the quasar rest frame. The wavelength corresponding to the quasar light echo surface, $\lambda_{\rm le}$, is described for various line-of-sight angles: $\theta ['']= $ 20 (black), 100 (blue), 400 (green), and 1000 (red). Different types of lines show the results for quasars at different redshifts: $z_{q} = 1.5$ (dotted lines), 2.5 (solid lines), and 5.0 (dot-dashed lines). The horizontal dashed line highlights the Ly$\alpha$ wavelength at the quasar rest frame.
}
\end{figure}

\begin{figure*}
\begin{tabular}{cc}
\begin{minipage}{0.5\hsize}
\begin{center}
\includegraphics[width=\linewidth]{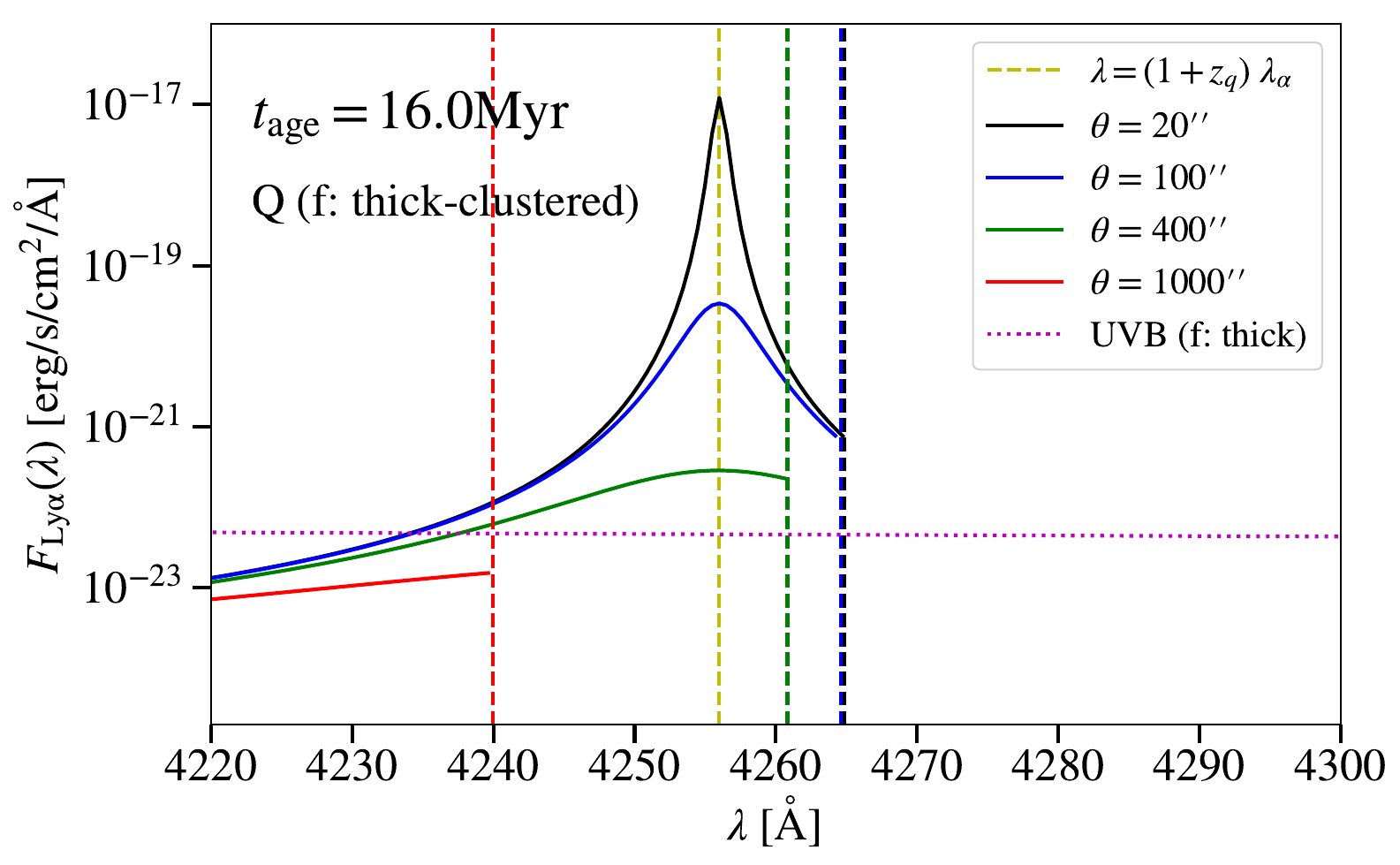}
\end{center}
\end{minipage} 
\begin{minipage}{0.5\hsize}
\begin{center}
\includegraphics[width=\linewidth]{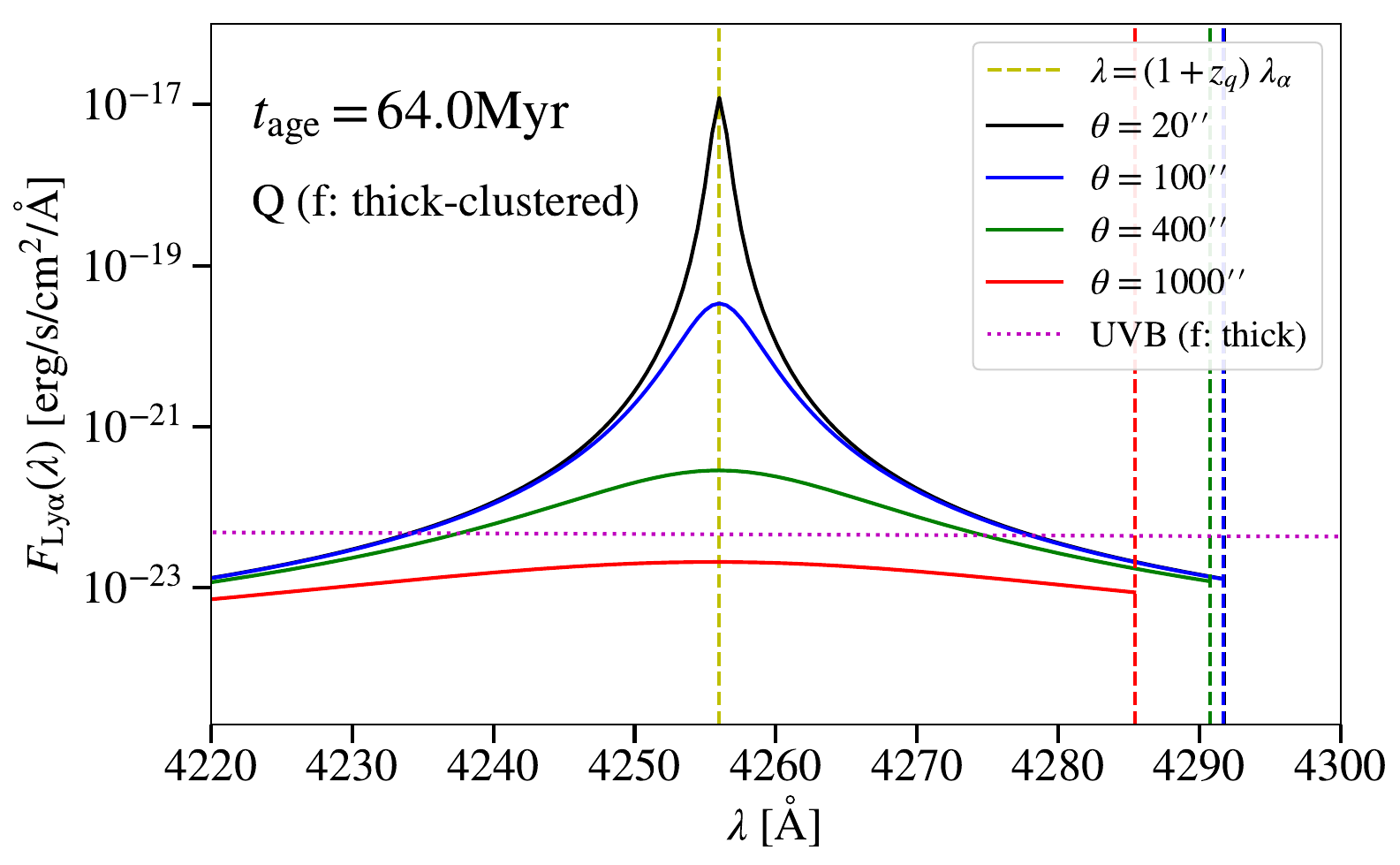}
\end{center}
\end{minipage}
\end{tabular}
\caption{\label{fig:flux_all_theta} Fluorescence Ly$\alpha$ emission flux from LLSs induced by a quasar at $z_{q}=2.5$. The two panels correspond to quasars with different ages: $t_{\rm age} =$ 16.0 (left) and 64.0~Myr (right). In each panel, the fluorescence Ly$\alpha$ fluxes per fiber (equations~\ref{eq:j_lya_f_thick} and \ref{eq:phi_q_ll}), computed with the incidence for LLSs clustered around quasars, $l_{q}^{\rm LLS}$, are described as solid lines for various line-of-sight angles in the same manner as Fig.~\ref{fig:lambda_le}. The horizontal axis, $\lambda$, is the observed wavelength of Ly$\alpha$ photons and the vertical dashed lines are the upper limits, $\lambda_{\rm le}(z_{q}, t_{\rm age}, \theta)$, (the same color for each line-of-sight angle) and the Ly$\alpha$ wavelength at the quasar redshift (yellow). We also show the contribution of the Ly$\alpha$ fluorescence from LLSs in the IGM (i.e., with the incidence of $l_{\rm IGM}^{\rm LLS}$) due to the UV background, with a horizontal dotted line.}
\end{figure*}

\section{Quasar-induced Lyman-alpha emission observed by spectroscopic fibers} \label{sec:quasar-induced_Lya}

Based on the expressions of the scattering and fluorescence in the previous two sections, in this section, we formulate the contribution of the quasar-induced Ly$\alpha$ emission to the photon flux observed by spectroscopic fibers.

\subsection{Observed flux of the quasar-induced Ly$\alpha$ photons\label{sec:F_lya}}

In general, the surface brightness of Ly$\alpha$ emissions from the IGM at an arbitrary position, over a finite, line-of-sight (physical) length is
\begin{eqnarray}     
     \mathrm{SB}_{\rm Ly\alpha} = \frac{1}{(1+z)^4}\int j_{\rm Ly\alpha} \mathrm{d}s \label{eq:I_lya}
\end{eqnarray}
(in $\rm{erg}\ \rm{s}^{-1}\ \rm{cm}^{-2}\ \rm{arcsec}^{-2}$), where we account for the reduction due to cosmological distances with the factor of $1/(1+z)^4 (=d_{A}^2/d_{L}^2)$. When we observe photons through a spectroscopic fiber with an aperture, a solid angle of $\Omega_{\rm f}$, each pixel as a wavelength bin corresponds to a length scale along the line of sight. A variation in the wavelength of observed Ly$\alpha$ photons due to the expansion of the Universe is related to a line-of-sight (physical) length $\mathrm{d}s$ as follows:
\begin{eqnarray}
    \mathrm{d}s &=& c\mathrm{d}t 
    \nonumber \\
    &=& \left.\frac{c}{H}\frac{\mathrm{d}a}{a}\right|_{z=z_{r}} = - \frac{c}{H(z_{r})}\frac{\mathrm{d}\lambda_{\alpha, {\rm obs}}}{\lambda_{\alpha, {\rm obs}}},  \label{eq:dR}
\end{eqnarray}
where $\lambda_{\alpha, {\rm obs}} = (1+z_{r})\lambda_{\alpha}$ is the observed wavelength of Ly$\alpha$ photons scattered at the point $R$. This means that Ly$\alpha$ photons detected in the wavelength range $\lambda_{\alpha, {\rm obs}}\sim\lambda_{\alpha, {\rm obs}}+\mathrm{d}\lambda_{\alpha, {\rm obs}}$ were scattered by gas clouds (i.e., the neutral hydrogen atoms in them) within a cylinder of $A_{\rm f} \times \mathrm{d}s$ located at the point $R$ (see Fig.~\ref{fig:around_R}), where $A_{\rm f} = d_{A}^2(z_{r})\Omega_{\rm f}$. From equations~(\ref{eq:I_lya}) and (\ref{eq:dR}), we obtain the following expression for Ly$\alpha$ flux per unit wavelength (in $\rm{erg}\ \rm{cm}^{-2}\ \rm{s}^{-1}\ \text{\AA}^{-1}$):
\begin{eqnarray}
    F_{\rm Ly\alpha}(\lambda) 
        &=& \Omega_{\rm f} \frac{\mathrm{d}\mathrm{SB}_{\rm Ly\alpha}}{\mathrm{d}\lambda} \left(\equiv \Omega_{\rm f}\ I_{\rm Ly\alpha}(\lambda) \right)
        \nonumber \\
        &=& \Omega_{\rm f}\left[\frac{j_{\rm Ly\alpha} }{(1+z_{r})^4} \frac{c}{H \lambda}\right]_{z_{r}=z_{r}(\lambda)}
        \label{eq:F_lya} 
\end{eqnarray}
Here $\lambda$ corresponds to the observed wavelength of Ly$\alpha$ photons, i.e., $\lambda_{\alpha, {\rm obs}}$ in equation~(\ref{eq:dR}). We emphasize that this flux corresponds to Ly$\alpha$ photons observed through a single fiber with the solid angle $\Omega_{\rm f}$.

\subsection{Results\label{sec:results}}

Before presenting the results of the quasar-induced Ly$\alpha$ emission flux, let us introduce an important concept for considering quasar age. The redshift of the reflection point $R$, $z_{r}$, has the upper limit corresponding to the light echo surface, $z_{r,{\rm le}}(z_{q}, t_{\rm age}, \theta)$, for each line-of-sight direction (see Section~\ref{sec:light_echo}). We then define the upper limit of the wavelength as $\lambda_{\rm le}(z_{q}, t_{\rm age}, \theta) \equiv (1+z_{r,{\rm le}})\lambda_{\alpha}$. Fig.~\ref{fig:lambda_le} shows the relation between the upper limit of the wavelength and the quasar age, and its redshift dependence. We see that the upper limit shifts to the longer wavelength side as the quasar becomes older because the light echo surface of older quasars is located further from the quasars (see Fig.~\ref{fig:LEs}). For a fixed quasar age, the upper limit for the inner line of sight has a longer value, reflecting the {\it curve} of the light echo surface. On the other hand, at each redshift, the upper limits for all lines of sight converge when the quasar age is old enough that the curvature of the light echo surface can be ignored. Equation~(\ref{eq:chi_le}) suggests that for old quasars (or small line-of-sight separations), the position of the light echo surface relative to the quasar in the comoving frame is roughly determined by the quasar age in the conformal time, $\Delta \eta_{\rm age}$. Therefore, the slope of each line is approximated by the Hubble parameter at the quasar redshift, $H(z_{q})$, resulting in steep slopes for quasars at higher redshifts.

First, let us look at the characteristic features of the quasar-induced Ly$\alpha$ emission observed by spectroscopic fibers. The solid lines in the two panels of Fig.~\ref{fig:flux_all_theta} demonstrate the fluorescence Ly$\alpha$ fluxes from optically thick gas clouds (i.e., LLSs) induced by different quasar ages: $t_{\rm age} =$ 16.0 (left) and 64.0~Myr (right). The flux for each line-of-sight separation does not change between different quasar ages except for the upper limit of the wavelength $\lambda_{\rm le}$ (vertical dashed lines) because the flux does not depend on the quasar age while the upper limit of the wavelength does. We see that the flux has a larger value when the reflection point $R$ is located closer to the quasar, i.e., at a radial distance closer to $\chi_{q}$ (yellow vertical dashed line) or in a direction of smaller $\theta$, since the quasar-induced Ly$\alpha$ emission flux is proportional to the quasar flux $F_{q}$. 

\begin{figure}
  \begin{center}
   \includegraphics[width=\linewidth]{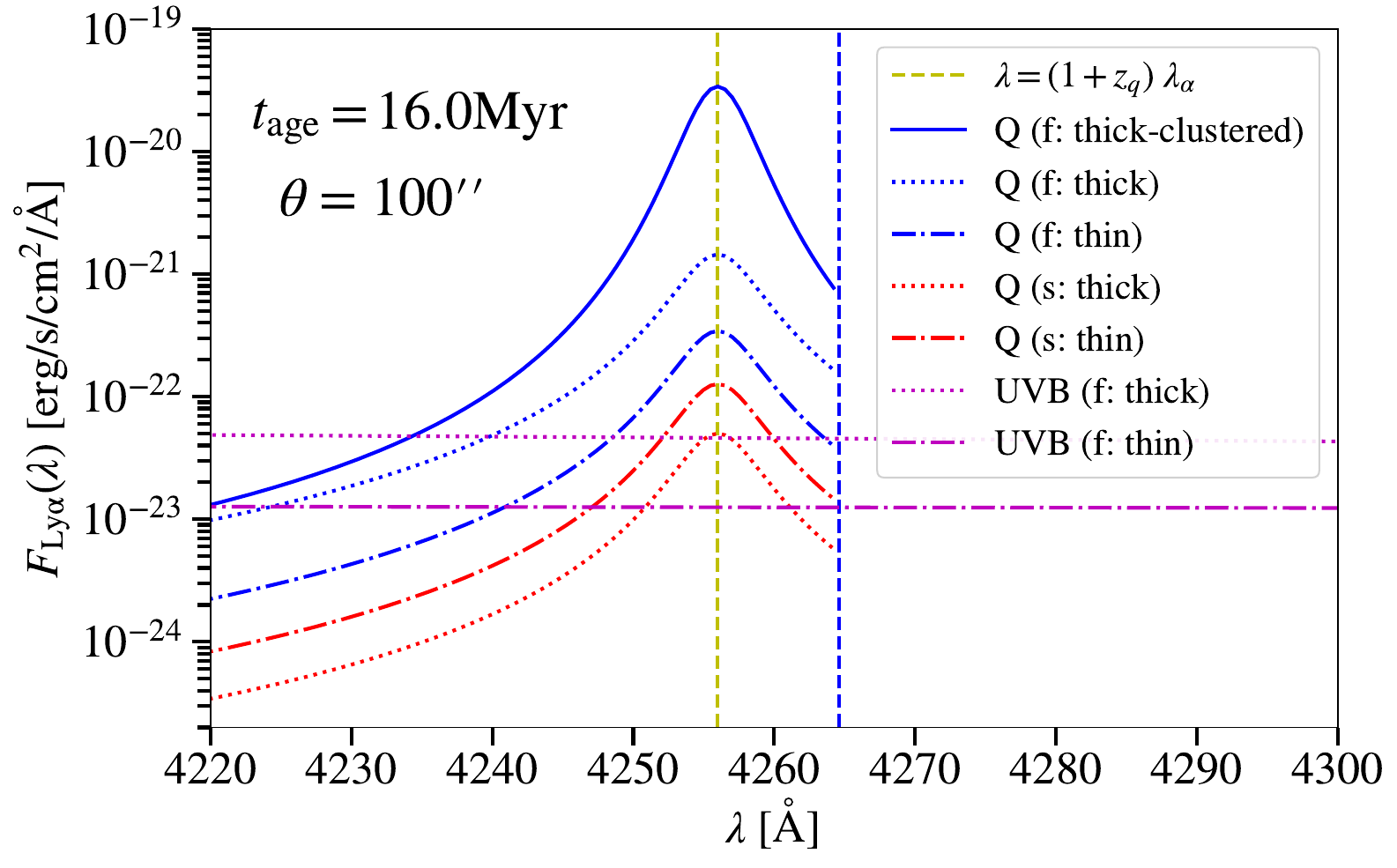}
  \end{center}
  \caption{\label{fig:flux_all_type} Comparison of different types of Ly$\alpha$ emission from the IGM induced by a quasar of $t_{\rm age}=16.0$~Myr at $z=2.5$ and the UV background. Three colors correspond to the Ly$\alpha$ fluxes observed by a spectroscopic fiber of a line-of-sight angle $\theta = 100''$, due to the resonant scattering of quasar Ly$\alpha$ photons (red), the fluorescence of quasar ionizing photons (blue), and the fluorescence of the UV background (magenta). The dot-dashed lines represent the contributions of optically thin gas clouds, and the solid and dotted lines represent those of optically thick gas clouds with and without considering the clustering enhancement around the quasar (i.e., the incidences of $l_{q}$ and $l_{\rm IGM}$), respectively.
}
\end{figure}

\begin{figure}
  \begin{center}
   \includegraphics[width=\linewidth]{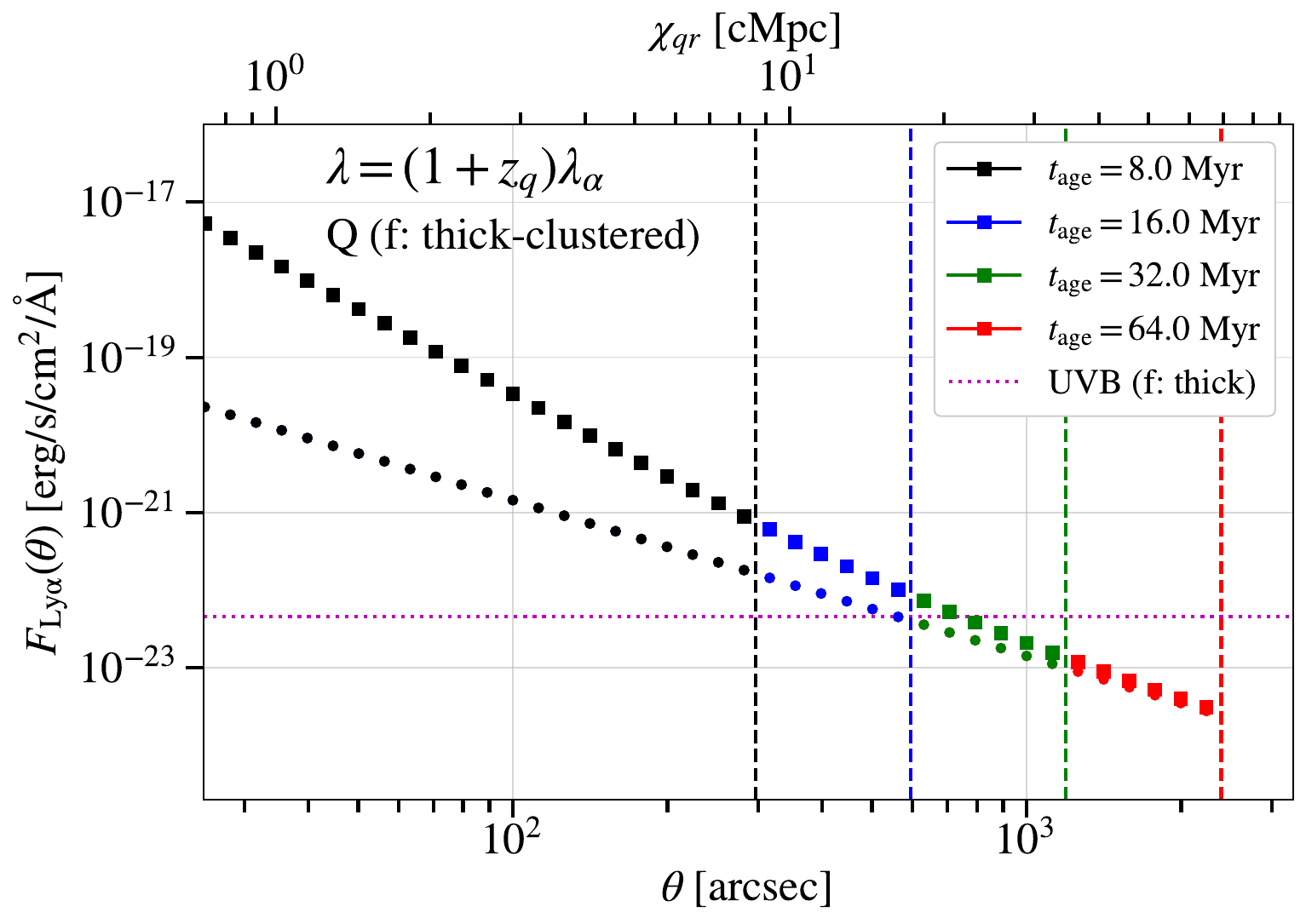}
  \end{center}
  \caption{\label{fig:flux_the} Angular dependence of the fluorescence Ly$\alpha$ flux. The fluorescence Ly$\alpha$ flux  from LLSs at a fixed wavelength of $\lambda = (1+z_{q})\lambda_{\alpha}$ (i.e., $z = z_{q}$) is plotted as the function of the line-of-sight separation, $\theta$, in both cases with $l^{\rm LLS}_{q}$ (squares) and $l^{\rm LLS}_{\rm IGM}$ (dots). The four colors correspond to quasars ($z_{q}=2.5$) with different ages: $t_{\rm age} =$ 8.0 (black), 16.0 (blue), 32.0 (green), and 64.0 Myr (red), overlapping for smaller line-of-sight separations. The vertical dashed lines are the maximum values of the angle, $\theta_{\rm max}(z_{q}, t_{\rm age})$, which satisfy the relation: $\chi_{r,{\rm le}}(\theta_{\rm max}, z_{q}, t_{\rm age}) = \chi_{q}$. We also show the comoving distance between the quasar and the scattering point, $\chi_{qr}$, on the upper horizontal axis, corresponding to $\theta$. The magenta dotted line represents the fluorescence from LLSs due to the UV background.}
\end{figure}

In Fig.~\ref{fig:flux_all_type}, we compare different processes of producing Ly$\alpha$ emission: scattering of quasar Ly$\alpha$ photons (in red), fluorescence of quasar ionizing photons (in blue), and fluorescence of the UV background (in magenta), for each gas cloud classification: optically thin (dot-dashed lines) and thick (dotted lines). While the contribution of optically thin clouds for resonant scattering is larger than that of optically thick clouds, this relationship is reversed for fluorescence. The most dominant contribution comes from the fluorescence of optically thick clouds (i.e., LLSs), which is larger than the resonant scattering of optically thick clouds (i.e., the LAF) by a factor of $\sim30$. This dominance of fluorescence in contributions from optically thick clouds can be understood as follows. Both Ly$\alpha$ emissivities produced through scattering (equation~\ref{eq:j_lya_s_thick}) and fluorescence (equation~\ref{eq:j_lya_f_thick}) are described by the product of the incidence of clouds and the number flux of quasar photons. The ratio of the number fluxes of quasar ionizing and Ly$\alpha$ photons is estimated as $\eta_{\rm thick}\Phi_{q,{\rm LL}}/\Phi_{q,{\rm Ly\alpha}}\simeq (\eta_{\rm thick}L_{q, \nu_{\rm LL}}/|\alpha|)/(L_{q}(\nu_{\alpha})W)\sim 4 \times 10^{3}$, which reflects the narrowness of the dimensionless EW for Ly$\alpha$ resonant scattering. On the other hand, the incidences for LLSs and the LAF differ by $l_{\rm IGM}^{\rm LLS}/l_{\rm IGM}^{\rm LAF}\sim 8\times 10^{-3}$ (see Fig.~\ref{fig:incidence_thick}). Combining these two ratios gives us the total factor of $\sim30$. In Section~\ref{sec:j_Lya}, we noted that the sum of $(\mathrm{d}f_{\rm C}/\mathrm{d}s) N_{\rm HI, c}$ for optically thin clouds (see equations~\ref{eq:def_thin_LAF} and \ref{eq:def_thin_LLS}) is overestimated because optically thick clouds are reluctantly included at the high column density end. However, we finally found that both the scattering and fluorescence contributions from optically thin gas clouds are smaller than the dominant contribution, i.e., fluorescence from optically thick gas clouds. In addition, we show the fluorescence Ly$\alpha$ emission from LLSs clustered around the quasar with a blue solid line in Fig.~\ref{fig:flux_all_type}, which is enhanced near the quasar compared with the case without considering the clustering effect. This feature can be seen more clearly in Fig.~\ref{fig:flux_the}, where the fluorescence Ly$\alpha$ fluxes from LLSs at a fixed wavelength, $F_{\rm Ly\alpha}(\theta, \lambda = (1+z_{q})\lambda_{\alpha})$, are represented as a function of the line-of-sight separation $\theta$. Including the clustering effect (squares) amplifies the original Ly$\alpha$ flux (dots), reflecting the power-law nature of our clustering model (equation~\ref{eq:incidence_QSO}).

We also show the fluorescence Ly$\alpha$ fluxes due to the UV background with magenta horizontal lines in Figs.~\ref{fig:flux_all_theta} and \ref{fig:flux_the}. The contribution of optically thick clouds (dotted line) is superior to that of optically thin clouds (dot-dashed line) in Fig.~\ref{fig:flux_all_theta}, as is the fluorescence due to quasar ionizing photons. In Fig.~\ref{fig:flux_the}, we find that the UV background fluorescence is smaller than the quasar-induced one in the inner region of $\theta \lsim 7 \times 10^{2\ \prime \prime}$. This suggests that within a radius of $\sim 20\ \rm{cMpc}$ (comoving megaparsecs) from a quasar at $z_{q} \simeq 2.5$, the contribution of quasar-induced Ly$\alpha$ photons from the IGM to the total Ly$\alpha$ flux dominates that of the UV background fluorescence.

\section{Cross-correlation with quasars} \label{sec:cross-corr}

In the previous sections, we focused on the quasar-induced Ly$\alpha$ emission in the IGM around a local quasar, gave the expressions of the Ly$\alpha$ emissivity due to resonant scattering and fluorescence, and calculated the Ly$\alpha$ fluxes (per fiber) for various quasar ages in various line-of-sight separations. To look further into the Ly$\alpha$ emissions around quasars, we here consider the cross-correlation between a quasar and the surface brightness of the Ly$\alpha$ emissions in the vicinity.

\subsection{Excess of Ly$\alpha$ emission by quasar radiation \label{sec:Lya_exce}}

Before introducing the formalism of cross-correlation, we would like to mention the excess of Ly$\alpha$ emission in the vicinity of quasars. While we used the mean IGM distribution that is averaged at redshifts and at distances from local quasars (see equations~\ref{eq:def_thin_LAF} and \ref{eq:def_thin_LLS}), each line of sight penetrates the IGM with different density contrasts. Therefore, the Ly$\alpha$ flux observed in each line-of-sight direction (i.e., fiber) is different from each other, even on the same line-of-sight separation $\theta$. In general, the excess of Ly$\alpha$ emission is described as a function of the line-of-sight direction, $\hat{\bf n}$:
\begin{eqnarray}
     \Delta I_{{\rm Ly}\alpha}(\hat{\bf n}, \lambda) 
     &=& \left(I_{{\rm Q}(s)} + I_{{\rm Q}(f)} + I_{{\rm UVB}(f)}\right)[n_{\rm c}(\hat{\bf n}, z_{r})] 
     \nonumber \\
     && \qquad \qquad - \ I_{{\rm UVB}(f)}[\bar{n}_{\rm c}(z_{r})]
     \nonumber \\
     && \qquad \qquad + \ \Delta I^{\rm SFG}_{{\rm Ly}\alpha}(\hat{\bf n}, z_{r}),  \label{eq:Delta_I}
\end{eqnarray}
where $z_{r}(\lambda) = \lambda/\lambda_{\alpha}-1$ and $I_{X}$ is the surface brightness per unit wavelength (in $\rm{erg}\ \rm{cm}^{-2} \ \rm{s}^{-1}\ \text{\AA}^{-1}\ \rm{arcsec}^{-2}$, see equation~\ref{eq:F_lya}). Here, the square bracket $[\cdots]$ specifies the type of the gas cloud number density that is used to compute the surface brightness and $\bar{n}_{\rm c}$ represents the mean number density, averaged at each redshift. While the first two lines are caused by HI atoms in the IGM, the last term, $\Delta I^{\rm SFG}_{{\rm Ly}\alpha}$, corresponds to Ly$\alpha$ photons emitted from star-forming galaxies, which are clustered around quasars (see Section~\ref{sec:SFG}).

Based on the discussions in Section~\ref{sec:results}, we can conclude that although LLSs are rarer than the LAF, they effectively reflect quasar ionizing photons, which are much more than the UV background or quasar Ly$\alpha$ photons. The above equation is then well approximated by 
\begin{eqnarray}
     \Delta I_{{\rm Ly}\alpha}(\hat{\bf n}, \lambda) 
     &\simeq& I^{\rm thick}_{{\rm Q}(f)}[n_{\rm c}(\hat{\bf n}, z_{r})] +\Delta I^{\rm SFG}_{{\rm Ly}\alpha}(\hat{\bf n}, z_{r}).  \label{eq:Delta_I_app}
\end{eqnarray}
In the following sections, aiming to estimate the excess of Ly$\alpha$ emission around quasars, we use this approximation. Note that in Fig.~\ref{fig:flux_all_type}, we considered the clustering enhancement only for the fluorescence from LLSs. Still, the enhancements for other processes might be different since their distance dependences (i.e., $r_{0}$ or $\gamma$ in the cross-correlation function, $\xi_{\rm QA}$) depend on the HI column density of gas clouds.

\subsection{Cross-correlation between quasar and Ly$\alpha$ emission}

In ongoing and future spectroscopic surveys, we will obtain a set of spectra in various directions around a large number of quasars. In this paper, we adopt the formalism used in previous studies by \citet{2016MNRAS.457.3541C, 2018MNRAS.481.1320C} and then define the quasar-Ly$\alpha$ emission cross-correlation for the $i$-th separation distance bin, $\chi_{qr, i}$, as the average of the residual surface brightness over $N_{i}$ pixels within the spherical shell of the $i$-th distance bin:
\begin{eqnarray}
     \xi_{{\rm q}\alpha, i} = \frac{1}{\sum_{j=1}^{N_{i}} w_{j}} \sum_{j=1}^{N_{i}}w_{j} \Delta I_{{\rm obs}, j}.  \label{eq:xi_qa}
\end{eqnarray}
Here $\Delta I_{{\rm obs}, j} \equiv I_{{\rm obs}, j} - \langle I_{{\rm obs}}(z_q) \rangle$ is the residual surface brightness at the $j$-th pixel, where $\langle I_{{\rm obs}}(z_q) \rangle$ represents the average of the surface brightness at $z=z_{q}$. The surface brightness in each pixel is weighed by the inverse variance of the flux, $w_{j}$.

Let us rewrite the above expression in terms of the excess of Ly$\alpha$ emission by assuming
\begin{eqnarray}
     \Delta I_{{\rm obs}}(\hat{\bf n}_{j}, \lambda_{j})
     = \Delta I_{{\rm Ly}\alpha}(\hat{\bf n}_{j}, \lambda_{j}) + \delta n(\hat{\bf n}_{j}, \lambda_{j}), \label{eq:Delta_I_mod}
\end{eqnarray}
where $(\hat{\bf n}_{j}, \lambda_{j})$ specifies the spatial position of the $j$-th pixel and $\delta n$ is the background noise per pixel with a zero mean and a variance of $\sigma_{n}(\lambda_{j})^{2}$. Although the variance of the flux is the sum of the intrinsic variance due to large-scale structure ($\Delta I_{{\rm Ly}\alpha}$ is a function of the HI number density) and the instrumental noise contribution, the large-scale structure variance is negligible \citep[e.g.,][]{2018MNRAS.481.1320C}. In the following, we then only consider the noise contribution to the flux variance, i.e., $w_{j} = 1/\sigma_{n}(\lambda_{j})^{2}$, which depends on only the wavelength, not the direction. Since the comoving distance between a quasar and an arbitrary emission position (point $R$: $z=z_{r}$), $\chi_{qr}$, is determined by both the direction angle $\theta$ and the observed Ly$\alpha$ wavelength $\lambda = (1+z_{r})\lambda_{\alpha}$, there are different pairs of $\theta$ and $\lambda$ for a single value of $\chi_{qr}$. We can then rewrite equation~(\ref{eq:xi_qa}) with the excess of Ly$\alpha$ emission by dividing $N_{i}$ pixels of $i$-th distance bin into $\theta$ bins:
\begin{eqnarray}
     \xi_{{\rm q}\alpha, i}
     &=& \frac{1}{\sum_{\theta_{m}}\sum_{j=1}^{N_{i}(\theta_{m})} w_{j}} \sum_{\theta_{m}} \sum_{j=1}^{N_{i}(\theta_{m})}w_{j} \Delta I_{\rm obs}(\hat{\bf n}_{j}, \lambda_{j})
     \nonumber \\
     &=& \frac{1}{\sum_{\theta_{m}}N_{\theta_{m}}\sum_{k=1}^{p_{i}(\theta_{m})} w_{k}} 
     \nonumber \\
     && \times \sum_{\theta_{m}} N_{\theta_{m}} \sum_{k=1}^{p_{i}(\theta_{m})}w_{k} \ \overline{\Delta I}_{\rm obs}(\theta_{m}, \lambda_{k}).
     \label{eq:xi_qa_re}
\end{eqnarray}
Here
\begin{eqnarray}
     \overline{\Delta I}_{\rm obs}(\theta_{m}, \lambda_{k})
     \equiv \frac{1}{N_{\theta_{m}}} \sum_{\substack{\hat{\bf n}_{j} \in {\rm bin} \ \theta_{m} \\ \lambda_{j} = \lambda_{k}}} \Delta I_{\rm obs}(\hat{\bf n}_{j}, \lambda_{j}), \label{eq:Delta_I_mod_ave}
\end{eqnarray}
where $\theta_{m}$ is the $m$-th direction angle bin and $N_{\theta_{m}}$ is the number of spectra in bin $\theta_{m}$. $N_{i}(\theta_{m})$ corresponds to the number of pixels that are placed on spectra in bin $\theta_{m}$, i.e., $\sum_{\theta_{m}}N_{i}(\theta_{m}) = N_{i}$. In addition, $p_{i}(\theta_{m})$ is the number of pixels on a spectrum of $\theta_{m}$ that are included within the spherical shell of $\chi_{qr, i}$, which satisfies the relation: $N_{i}(\theta_{m}) = N_{\theta_{m}}p_{i}(\theta_{m})$. 

The expression of equation~(\ref{eq:xi_qa_re}) shows that the quasar-Ly$\alpha$ emission cross-correlation is described as the sum of the Ly$\alpha$ emission excesses averaged over pixels within the rings of $\theta_{m}$ at radial distances $\lambda_{k}$, where the relation between $\theta_{m}$ and $\lambda_{k}$ is decided by $\chi_{qr, i}$. We can thus estimate the mean of $\xi_{{\rm q}\alpha}$, $\langle\xi_{{\rm q}\alpha}\rangle$, from the mean of the averaged emission excess, equation~(\ref{eq:Delta_I_mod_ave}). Using equations~(\ref{eq:Delta_I_app}) and (\ref{eq:Delta_I_mod}), the mean of $\overline{\Delta I}_{\rm obs}$ is evaluated by
\begin{eqnarray}
     &&\left\langle \overline{\Delta I}_{\rm obs}(\theta_{m}, \lambda_{k}) \right\rangle
     \nonumber \\
     &=& \left\langle I^{\rm thick}_{{\rm Q}(f)}[n_{\rm c}(\hat{\bf n}, z_{r}(\lambda_{k}))] + \Delta I^{\rm SFG}_{{\rm Ly}\alpha}(\hat{\bf n}, z_{r}(\lambda_{k})) \right\rangle_{\hat{\bf n} \in {\rm bin} \ \theta_{m}}
     \nonumber \\
     &=& I^{\rm thick}_{{\rm Q}(f)}\left[\bar{n}_{\rm c}\left(\theta_{m}, z_{r}(\lambda_{k})\right)\right] + \Delta I^{\rm SFG}_{{\rm Ly}\alpha}(\theta_{m}, z_{r}(\lambda_{k})), \label{eq:Delta_I_mod_ave_mean}
\end{eqnarray}
where $\bar{n}_{\rm c}(\theta_{m}, z_{r})$ is the mean number density for given direction angle and redshift. The first term represents the contribution of the quasar-induced Ly$\alpha$ emission, $\xi^{\rm Q}_{{\rm q}\alpha}$, which can be calculated by using the formulation described in Section~\ref{sec:fluorescence} and recalling that $I^{\rm thick}_{{\rm Q}(f)}$ is not zero if $\lambda_{k}<\lambda_{\rm le}(z_{q}, t_{\rm age}, \theta_{m})$. The contribution due to the second term corresponds to equation~(\ref{eq:xi_SFG}), $\xi^{\rm SFG}_{{\rm q}\alpha}$, with $s=\chi_{qr}$. In the following, we will primarily focus on the first term, i.e., $\langle\xi_{{\rm q}\alpha}\rangle=\xi^{\rm Q}_{{\rm q}\alpha}$, which is the main interest in this paper, and compare it with the second term accordingly.

\begin{figure}
  \begin{center}
   \includegraphics[width=\linewidth]{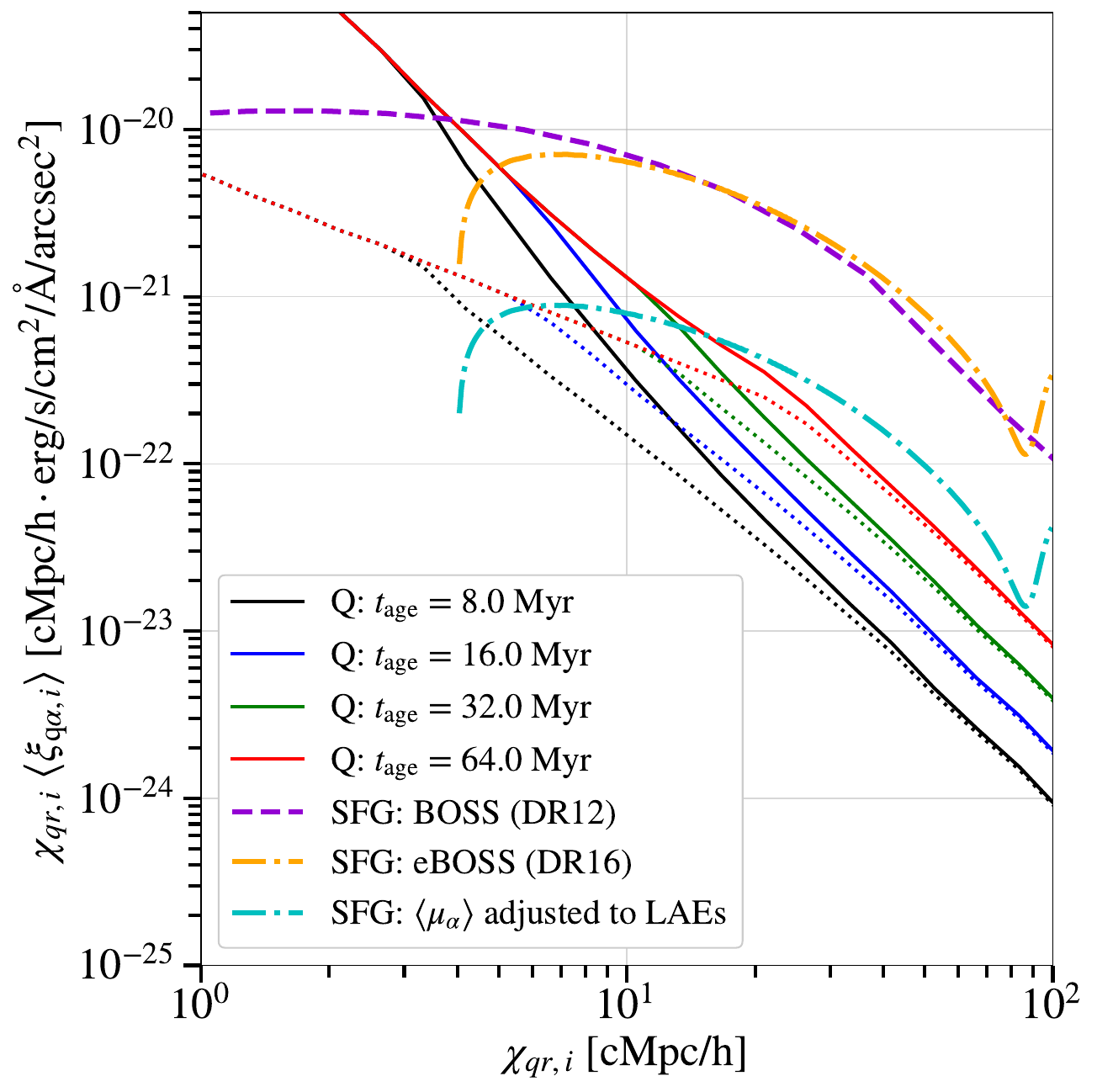}
  \end{center}
  \caption{\label{fig:x_corr} Quasar-induced contribution to the cross-correlation between a quasar at $z_{q}=2.5$ and the surrounding Ly$\alpha$ emission excess. The solid and dotted lines are the means of the cross-correlation, $\xi^{\rm Q}_{{\rm q}\alpha}$, computed with the incidences of $l_{q}$ and $l_{\rm IGM}$, respectively. The four colors correspond to quasars with different ages: $t_{\rm age} =$ 8.0 (black), 16.0 (blue), 32.0 (green), and 64.0 Myr (red). For comparison, we put models based on the cross-correlation between quasars and star-forming galaxies, $\xi^{\rm SFG}_{{\rm q}\alpha}$, which were fitted to the BOSS \citep[DR12,][]{2018MNRAS.481.1320C} and eBOSS \citep[DR16,][]{2022ApJS..262...38L} data, as the dashed (purple) and dot-dashed (orange) lines, respectively. The cyan dot-dashed line represents the same cross-correlation model as \citet{2022ApJS..262...38L}, but with $\langle \mu_{\alpha} \rangle$ adjusted to observations of LAEs and $\beta_{\alpha}$ derived by assuming $\Lambda$CDM. We will discuss the gap between our model and the measurements in Section~\ref{sec:potential_contributions}.}
\end{figure}

\subsection{Forecast for PFS \label{sec:forecast}}

We now calculate the cross-correlation between quasar and Ly$\alpha$ emission using the PFS cosmology survey parameters \citep{2014PASJ...66R...1T}. In this study, we calculate the background noise assumed in the PFS survey using {\it PFS Exposure Time Calculator and Spectrum Simulator},\footnote{\url{https://github.com/Subaru-PFS/spt_ExposureTimeCalculator}} which is based on the package developed by \citet{2012arXiv1204.5151H}. This exposure time calculator (ETC) allows us to compute the SNR of an object or the background noise in various conditions. In particular, we make use of the variance of the background noise per one exposure, $\sigma_{n, {\rm exp}}(\lambda)$ (in $\rm{erg}\ \rm{cm}^{-2}\ \rm{s}^{-1}\ \text{\AA}^{-1} \rm{arcsec}^{-2}$), with the observational conditions assumed in the PFS survey (e.g., the single exposure time: $t_{{\rm exp}} = 450$~seconds or the number of exposures: $N_{\rm exp} = 2$). Since the noise variance is suppressed by a factor of $1/\sqrt{N_{\rm exp}}$ when averaging the observed fluxes over all exposures, the variance of the background noise per pixel is estimated as $\sigma_{n}(\lambda) = \sigma_{n, {\rm exp}}(\lambda)/\sqrt{N_{\rm exp}}$. In addition, we set the width of the pixel (or wavelength) bin to $\lambda_{\alpha, {\rm obs}}/(\lambda/\Delta \lambda)_{\rm res} = 2.24$\AA, where we assumed the PFS spectral resolution: $(\lambda/\Delta \lambda)_{\rm res} = 1900$.

Fig.~\ref{fig:x_corr} shows the quasar-induced contribution to the cross-correlation for a quasar at $z_{q}=2.5$, computed with the width of the direction angle bin of $5^{\prime\prime}$.\footnote{While the number of spectra in bin $\theta_{m}$, $N_{\theta_{m}}$, generally scales as the sky area of the ring of $\theta_{m}$ bin and the number of spectra per unit sky area, the cross-correlation, equation~(\ref{eq:xi_qa_re}), only depends on the former because the latter is canceled out in the numerator and the denominator. Therefore, we here adopted a large enough value as the number of spectra so that each $N_{\theta_{m}}$ is larger than unity.} The solid (dotted) lines are the mean values, $\langle\xi_{{\rm q}\alpha, i}\rangle = \xi^{\rm Q}_{{\rm q}\alpha}$, computed from equations~(\ref{eq:xi_qa_re})-(\ref{eq:Delta_I_mod_ave_mean}) with the incidence of $l_{q}$ ($l_{\rm IGM}$), and each color represents a different quasar age. We can see that both cases, with and without the clustering enhancement, are consistent with the plots of Fig.~\ref{fig:flux_the} except for the factor of $\chi_{qr}$, showing that the cross-correlation actually corresponds to the average of the Ly$\alpha$ emission excess at the same distance from quasars. The main difference with Fig.~\ref{fig:flux_the} is the cross-correlation for younger quasars starting to decrease at a point closer to the quasar because the size of the quasar light echo surface is smaller (corresponding to the upper limit of the wavelength shifting to the shorter wavelength side in Fig.~\ref{fig:flux_all_theta}).

To compare with observations, we show the measurements of the BOSS \citep[][]{2018MNRAS.481.1320C} and eBOSS \citep[][]{2022ApJS..262...38L} data as the dotted and dashed lines, respectively. The mean cross-correlations, $\langle\xi_{{\rm q}\alpha, i}\rangle=\xi^{\rm SFG}_{{\rm q}\alpha}$, are their best-fitting model with its scale-dependence following the cross-correlation between quasars and star-forming galaxies, which was described in Section~\ref{sec:SFG}.\footnote{Note that \citet[][]{2018MNRAS.481.1320C} adopted the same model as equation~(\ref{eq:xi_SFG}), but without truncating small-scale nonlinear effects.} The mean redshifts of their quasar catalogs, $z=2.55$ for \citet[][]{2018MNRAS.481.1320C} and $z=2.40$ for \citet[][]{2022ApJS..262...38L}, are quite similar to ours, $z_{q} =2.5$. We can see that both measurements are consistent with each other. Comparing our model with the measurements, we found that the quasar-induced Ly$\alpha$ emission can account for $\sim10\%$ of the Ly$\alpha$ emission excess in the outer region of $\gsim10\ \rm{cMpc}\ \rm{h}^{-1}$. On the other hand, in the inner region of $\lsim10\ \rm{cMpc}\ \rm{h}^{-1}$, the case with $l_{q}$ is higher than the measurements while the case with $l_{\rm IGM}$ is lower. This implies that we need to revisit the cross-correlation model that we used to consider the clustering of gas clouds (i.e., equation~\ref{eq:incidence_QSO}). 

\citet[][]{2022ApJS..262...38L} found that the best-fitting values of $\langle \mu_{\alpha} \rangle$ and $\beta_{\alpha}$ are $\vpm{1.13}{0.57}{0.53} \times 10^{-21} \rm{erg}\ \rm{s}^{-1}\ \rm{cm}^{-2}\ \text{\AA}^{-1}\ \rm{arcsec}^{-2}$ and $\vpm{0.07}{1.65}{0.73}$, respectively, which determine the cross-correlation $\xi^{\rm SFG}_{{\rm q}\alpha}$ (the orange dot-dashed line in Fig.~\ref{fig:x_corr}). However, as they discussed in the paper, these numbers do not seem reasonable. For example, the corresponding comoving Ly$\alpha$ luminosity density, $\rho_{\alpha} = 4\pi\langle \mu_{\alpha} \rangle (H(z)/c)\lambda_{\alpha}(1+z)^{2}= 6.6 \times 10^{40}\rm{erg}\ \rm{s}^{-1}\ \rm{cMpc}^{-3}$, is roughly an order of magnitude higher than that inferred from the Ly$\alpha$ luminosity function of LAEs \citep[e.g., $\rho_{\alpha}=0.74 \times 10^{40}\rm{erg}\ \rm{s}^{-1}\ \rm{cMpc}^{-3}$ at $z=2.5$,][]{2018MNRAS.476.4725S}. Furthermore, assuming $\Lambda$CDM, $\beta_{\alpha}\simeq \Omega_{m}(z=2.40)^{0.55}/b_{\alpha} = 0.32$ with $b_{\alpha}=3$, which is slightly higher than their best-fit value although its uncertainty is quite large. The cyan dot-dashed line in Fig.~\ref{fig:x_corr} shows the same cross-correlation model as \citet[][]{2022ApJS..262...38L}, but with these two parameters set to the more reasonable values: (the corresponding) $\rho_{\alpha}=0.74 \times 10^{40}\rm{erg}\ \rm{s}^{-1}\ \rm{cMpc}^{-3}$ and $\beta_{\alpha} = \Omega_{m}(z=2.40)^{0.55}/b_{\alpha}$ with $b_{\alpha}=3$. The amplitude is an order of magnitude smaller than the eBOSS best-fit model and comparable with the quasar-induced contribution, $\xi^{\rm Q}_{{\rm q}\alpha}$, with $t_{\rm age} =$ 64 Myr on scales of $\sim10{\rm -}20\ \rm{cMpc}\ \rm{h}^{-1}$. This result appears to indicate that even if the quasar lifetime is relatively long, $t_{\rm age} \gsim$ 50 Myr, these two contributions of $\xi^{\rm Q}_{{\rm q}\alpha}+\xi^{\rm SFG}_{{\rm q}\alpha}$ (based on equation~\ref{eq:Delta_I}) cannot explain the measurements. However, \citet[][]{2022ApJS..262...38L} argued that Ly$\alpha$ emission sources other than LAEs (e.g., UV-selected galaxies with small Ly$\alpha$ EW or extended Ly$\alpha$ halos outside a typical aperture) could increase the Ly$\alpha$ luminosity density although the result changes by a factor of 4 at most, depending on the assumed UV luminosity function model. We then conclude that our model for the quasar-induced Ly$\alpha$ emission, combined with the contribution of star-forming galaxies, is not in conflict with these previous measurements. In Section~\ref{sec:potential_contributions}, we will discuss the validity of the absorber clustering model and other potential contributions to the Ly$\alpha$ emission excess around quasars. 

Next, we consider the signal-to-noise ratio (SNR) for the cross-correlation to estimate the detection probability in the PFS survey. The SNR is defined by $\langle\xi_{{\rm q}\alpha, i}\rangle/\sigma_{\xi_{{\rm q}\alpha, i}}$, where the variance, $\sigma_{\xi_{{\rm q}\alpha, i}}$, is computed from equation~(\ref{eq:xi_qa_re}):
\begin{eqnarray}
     \sigma_{\xi_{{\rm q}\alpha, i}}^{2} &\equiv& \left\langle \left(\xi_{{\rm q}\alpha, i} - \left\langle \xi_{{\rm q}\alpha, i} \right\rangle \right)^{2} \right\rangle
     \nonumber \\
     &=& \frac{1}{\left(\sum_{\theta_{m}} N_{\theta_{m}} \sum_{k=1}^{p_{i}(\theta_{m})} w_{k}\right)^{2}} 
     \sum_{\theta_{m}} N_{\theta_{m}} \sum_{k=1}^{p_{i}(\theta_{m})}w_{k}^{2} \sigma_{n}^{2}(\lambda_{k}),
     \nonumber \\
     &=& \left(\sum_{\theta_{m}} N_{\theta_{m}} \sum_{k=1}^{p_{i}(\theta_{m})} \frac{1}{\sigma_{n}^{2}(\lambda_{k})}\right)^{-1}.
     \label{eq:var_xi_qa}
\end{eqnarray}
Here we used the relation: $\left\langle \delta n(\hat{\bf n}_{j} , \lambda_{j}) \right\rangle^{2} = \sigma_{n}^{2}(\lambda_{j})$. Although we should focus on the cross-correlation for a single quasar if we seek to investigate its activity, we here consider stacking all spectra around a number of quasars to make the SNR higher as much as possible at the risk of smoothing individual quasar spectra. Since the number of spectra in bin $\theta_{m}$, $N_{\theta_{m}}$, is determined by the sky area of the ring of $\theta_{m}$ bin and the number of spectra per unit sky area, $n_{\rm fiber}$, we can then amplify the number of spectra with the number of quasars, $N_{q}$: $N_{\theta_{m}} \propto n_{\rm fiber}\ N_{q}$. 

In the PFS project, the Prime Focus Instrument has the number of fibers of $2400$ and the field of view area of $1.098 \ {\rm deg}^{2}$, and will survey over $1400 \ {\rm deg}^2$. We then set $n_{\rm fiber}$ to $2\times2400/1.098 = 4.4\times10^{3}\ {\rm deg}^{-2}$, considering two visits per each field scheduled. \citet{2018MNRAS.481.1320C} used the Data Release 12 (DR12) quasar catalog from the SDSS-III/BOSS survey, over an area of $\sim 9.4\times10^{3}\ {\rm deg}^{2}$ \citep{2017A&A...597A..79P} and selected $2.2\times 10^{5}$ quasars within the redshift range of $z=2.0{\rm -}3.5$. In this forecast, we assume that we make use of their selected quasar sample over the PFS survey area, which is completely overlapped with the BOSS survey area: $N_{q} = 2.2\times 10^{5}\times 1400/(9.4\times10^{3})=3.3\times10^{4}$.

\begin{figure}
\begin{center}
\includegraphics[width=\linewidth]{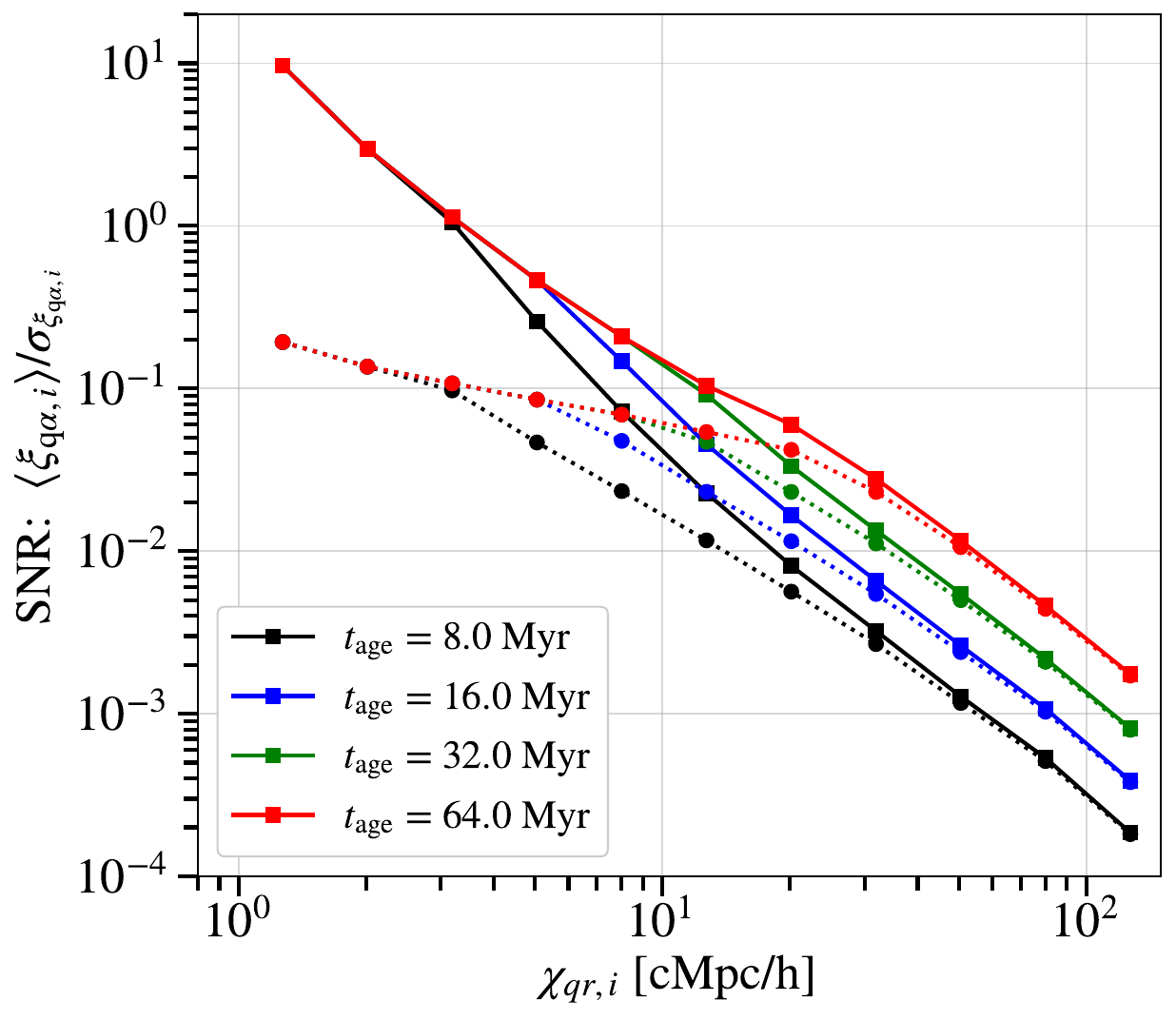}
\end{center}
\caption{\label{fig:SNR_PFS} Expected signal-to-noise ratio (SNR) for the quasar-induced Ly$\alpha$ emission in the PFS survey. The solid and dotted lines correspond to the ones in Fig.~\ref{fig:x_corr}. In all cases, the variance, $\sigma_{\xi_{{\rm q}\alpha, i}}$, is calculated from the background noise assumed in PFS (see equation~\ref{eq:var_xi_qa}) and each point on lines represents the center of each separation distance bin.}
\end{figure}

In Fig.~\ref{fig:SNR_PFS}, we show the signal-to-noise ratio (SNR) for the quasar-induced contribution, $\langle\xi_{{\rm q}\alpha, i}\rangle = \xi^{\rm Q}_{{\rm q}\alpha}$, expected in the PFS survey. The solid and dotted lines correspond to the quasar-induced contribution shown in Fig.~\ref{fig:x_corr}, except for adopting the width of $\theta_{m}$ bin of $15^{\prime\prime}$ in order to guarantee at least one fiber in each bin $\theta_{m}$. Note that when we compute the variance, $\sigma_{\xi_{{\rm q}\alpha, i}}$, for each quasar age, we assume that all quasars have the same age. Each point on lines represents the center of each separation distance bin, $\chi_{qr, i}$, and the interval is set to $\mathrm{d}\log_{10}\chi_{qr} = 0.2$. Since the variance depends on the distance from quasars as $\sigma_{\xi_{{\rm q}\alpha, i}}\propto\chi_{qr, i}^{-3/2}$,\footnote{As we will see in the next paragraph, $\sigma_{\xi_{{\rm q}\alpha, i}}^{2}$ is inversely proportional to the number of pixels within the $i$-th distance bin, $N_{i}$, which naively scale as the volume of the spherical shell of $\chi_{qr, i}$: $N_{i}\propto\chi_{qr, i}^{2}\mathrm{d}\chi_{qr, i} = \chi_{qr, i}^{3}\mathrm{d} \ln \chi_{qr, i}$.} the dependence of the SNR on the distance is similar to Fig.~\ref{fig:x_corr}: both SNRs for the incidences with $l_{q}$ and $l_{\rm IGM}$ decrease monotonically with the distance. We find that the SNR for quasars with an age older than 10~Myr is higher than $10^{-2}$ at distances of $\lsim20\ \rm{cMpc}\ \rm{h}^{-1}$. Note that the signals here include only the contribution due to quasar radiation, not star-forming galaxies, and then this does not mean that we cannot detect the cross-correlation signal in the PFS survey.

Finally, we evaluate how the SNR depends on the survey parameters. Since the variance of the background noise per pixel does not rapidly change within a radius of $\sim100\ \rm{cMpc}\ \rm{h}^{-1}$ from a quasar, we can roughly estimate the cross-correlation and its variance in simpler forms: $\xi_{{\rm q}\alpha, i} \simeq \sum_{j=1}^{N_{i}} \Delta I_{{\rm obs}, j} / N_{i}$; $\sigma_{\xi_{{\rm q}\alpha, i}}^{2} \simeq \sigma_{n}^{2}(\lambda_{q}) / N_{i}$, where $\lambda_{q} = (1+z_{q})\lambda_{\alpha}$ is the wavelength of the quasar Ly$\alpha$ photons. Hence, the SNRs in Fig.~\ref{fig:SNR_PFS} changes with the survey parameters as follows:
\begin{eqnarray}
     \textrm{S/N} 
     &\propto& \left( \frac{\sigma_{n}(\lambda_{q})}{9.2\times10^{-18}\rm{erg}\ \rm{cm}^{-2} \ \rm{s}^{-1}\ \text{\AA}^{-1}\ \rm{arcsec}^{-2}} \right)^{-1}
     \nonumber \\
     &\ & \times \left[\left( \frac{t_{\rm exp(tot)}}{900 \ \rm{s} } \right) \left( \frac{\lambda/\Delta\lambda}{1900} \right) \right.
     \nonumber \\
     &&\left. \qquad \quad \left( \frac{N_{q}}{3.3\times10^{4}} \right) \left( \frac{n_{\rm fiber}}{4.4\times10^{3}\ \rm{deg}^{-2}} \right)  \right]^{1/2},
\end{eqnarray}
where $t_{\rm exp(tot)} \equiv N_{\rm exp}t_{\rm exp}$ is the total exposure time. From the above expression, it is clear that we can improve the SNR by increasing the exposure time and the number of pixels or suppressing the background noise per pixel. However, we would like to note that constraining the quasar age using the quasar-Ly$\alpha$ emission cross-correlation is still challenging because we need to distinguish the quasar-induced Ly$\alpha$ emission from star-forming galaxies or the unknown residual contribution.

\section{Discussions} \label{sec:discussion}

In this study, we expected to stack a number of spectroscopic fibers around quasars in ongoing redshift surveys, estimated the flux of Ly$\alpha$ emission from the IGM that is induced by the quasar illumination, and predicted the detectability of the cross-correlation between quasar and Ly$\alpha$ emission. We will discuss the limitation, efficiency, and redshift dependence in the following subsections.

\subsection{Potential contributions to the Ly$\alpha$ emission excess \label{sec:potential_contributions}}

\citet{2016MNRAS.457.3541C} discussed in detail possible contributions to the excess of Ly$\alpha$ emission around quasars when interpreting their measurement of the quasar-Ly$\alpha$ emission cross-correlation. In this subsection, we point out contributions that we have missed in this paper, focusing in particular on whether they are related to quasar illumination or not.

\subsubsection{Ly$\alpha$ broad emission line}

In this study, we used a quasar spectrum model constructed by stacking over observed quasar spectra \citep[][]{2015MNRAS.449.4204L} as typical fluxes of Ly$\alpha$ and ionizing photons radiated from quasars. The most obvious effect leading to the enhancement of the Ly$\alpha$ emission excess is the contribution from the Ly$\alpha$ broad emission line. Although we only considered the continuum component as quasar spectra, the width of the Ly$\alpha$ emission line is quite large, with an FWHM of $\sim 50\ $\AA \ in the rest frame \citep[e.g.,][]{2016MNRAS.457.3541C}, which corresponds to a distance of $\sim 50\ {\rm pMpc}$ at $z=2.5$. The scattered Ly$\alpha$ flux scales as the flux of quasar Ly$\alpha$ photons and is then boosted by the additional contribution from the broad emission line in the vicinity of quasars.

\subsubsection{Anisotropic quasar radiation}

In addition, the anisotropy of quasar radiation could change the shape of the quasar-Ly$\alpha$ emission cross-correlation. Unification models of AGN \citep[e.g.,][]{1993ARA&A..31..473A,2015ARA&A..53..365N} suggest that quasar radiation is emitted in a cone with an opening angle and obscured by a torus of gas and dust in the perpendicular direction. This anisotropic feature of quasar emission has been confirmed by several observations \citep[e.g.,][]{2004Natur.430..999W,2007ApJ...655..735H,2013ApJ...776..136P,2019ApJ...884..151J}. Thus, the quasar spectrum model that we used here, which is based on the radiation along the line of sight, might not be suitable for modeling the fluxes of quasar Ly$\alpha$ and ionizing photons that are emitted in the perpendicular direction. In other words, by analyzing the anisotropy of the quasar-Ly$\alpha$ emission cross-correlation, we could measure the inclination angle of the emission cone or deduce its relationship with the quasar type: unobscured quasars (type~1) or obscured quasars (type~2) \citep[e.g.,][]{2003AJ....126.2125Z}. 

This anisotropic radiation also affects the clustering model of HI gas clouds around quasars. The assumed model for the cross-correlation function between quasars and absorbers, $\xi_{\rm QA}(r) = (r/r_{0})^{-\gamma}$, is spherically symmetric and characterized by the correlation length $r_{0}$ and the power-law index $\gamma$. For these parameters, we assumed values obtained by measuring the Ly$\alpha$ absorption transverse to quasar sightlines using background sightlines of projected quasar pairs \citep[][]{2013ApJ...776..136P}. However, \citet{2007ApJ...655..735H} found that the clustering of absorbers in the transverse direction is more enhanced than that in the line-of-sight direction and concluded that this anisotropic clustering is caused by the anisotropic quasar emission, i.e., the deficiency of quasar ionizing photons in the transverse direction and the photoevaporation of absorbers in the line-of-sight direction. This conclusion is supported by a study of Ly$\alpha$ emission around quasars \citep{2013ApJ...766...58H}, where they did not detect Ly$\alpha$ emission at the location of optically thick absorbers along background quasar sightlines. Based on these discussions, the overprediction of our model in Fig.~\ref{fig:x_corr} is at least partly due to ignoring the following two effects driven by the anisotropic radiation: fewer quasar ionizing photons in the transverse direction and fewer LLSs in the line-of-sight direction.

In addition, as another type of anisotropy, we have to mention redshift-space distortions (RSD). Although our absorber clustering model, $\xi_{\rm QA}$, is spherically symmetric (i.e., monopole), the model parameters $r_{0}$ and $\gamma$ were determined through the distribution of Ly$\alpha$ absorbers in redshift space and then implicitly reflect the clustering enhancement due to RSD in $\xi^{\rm Q}_{{\rm q}\alpha}$, as well as $\xi^{\rm SFG}_{{\rm q}\alpha}$, which is defined based on the redshift-space dark matter correlation. However, in this paper, we have only focused on the monopole for both $\xi^{\rm Q}_{{\rm q}\alpha}$ and $\xi^{\rm SFG}_{{\rm q}\alpha}$, and considering higher multiples will help us distinguish between these two components, since the anisotropic feature due to RSD is clearly different from that caused by the quasar radiation cone or parabolic light echo.

\subsubsection{Contributions not related to quasar illumination}

The two contributions above increase/decrease the Ly$\alpha$ emission excess caused by quasar illumination and then depend on its luminosity or age. On the other hand, there are some effects that contribute to the Ly$\alpha$ excess but are not related to quasar radiation. One of them is a larger population of external radiation sources, e.g., star-forming galaxies, due to the denser environment around quasars and we took into account their contribution through the cross-correlation between quasars and star-forming galaxies. While external radiation sources clustered around quasars directly contribute to the local excess of the Ly$\alpha$ emission intensity, they also vary the quasar-induced Ly$\alpha$ emission from the IGM. In this study, we only considered Ly$\alpha$ and ionizing photons from local quasars as radiation sources illuminating the surrounding diffuse IGM through resonant scattering or fluorescence. However, total photons from external sources, e.g., star-forming galaxies, around quasars are locally enhanced and thus boost the IGM illumination \citep[e.g.,][]{2016ApJ...822...84M}.

Another possible physical process that produces Ly$\alpha$ emission is collisional excitation caused by HI atoms and free electrons. In quasar halos, gas radiates its gravitational potential energy (i.e., cooling radiation) through dissipation processes, including collisional excitation. This process has been used to explain extended Ly$\alpha$ emission around quasars (\citealp{2000ApJ...537L...5H,2001ApJ...562..605F}; see \citealp[e.g.,][]{2012MNRAS.423..344R} for previous studies). The Ly$\alpha$ production by the cooling radiation occurs in a local quasar halo and then contributes to the Ly$\alpha$ excess around the quasar \citep[e.g.,][]{2016MNRAS.457.3541C}. 

We note that it is not essential to precisely model the {\it intensity} of Ly$\alpha$ excess around quasars in terms of mapping the quasar light echo surface. As we saw in Fig.~\ref{fig:flux_all_theta}, the shape of the quasar light echo surface is geometrically determined by the quasar age and the speed of light, and can then be inferred from the upper limit of the wavelength rather than the intensity in itself. Therefore, even if we take into account the potential contributions discussed above, the information of the quasar light echo is not necessarily diluted, and in fact, the constraint on the quasar age can be tightened by constructing the model taking into account whether each contribution is caused by the quasar illumination or not.

\subsection{Attenuation and Efficiency \label{sec:attenuation}} 

\begin{figure}
\begin{center}
\includegraphics[width=\linewidth]{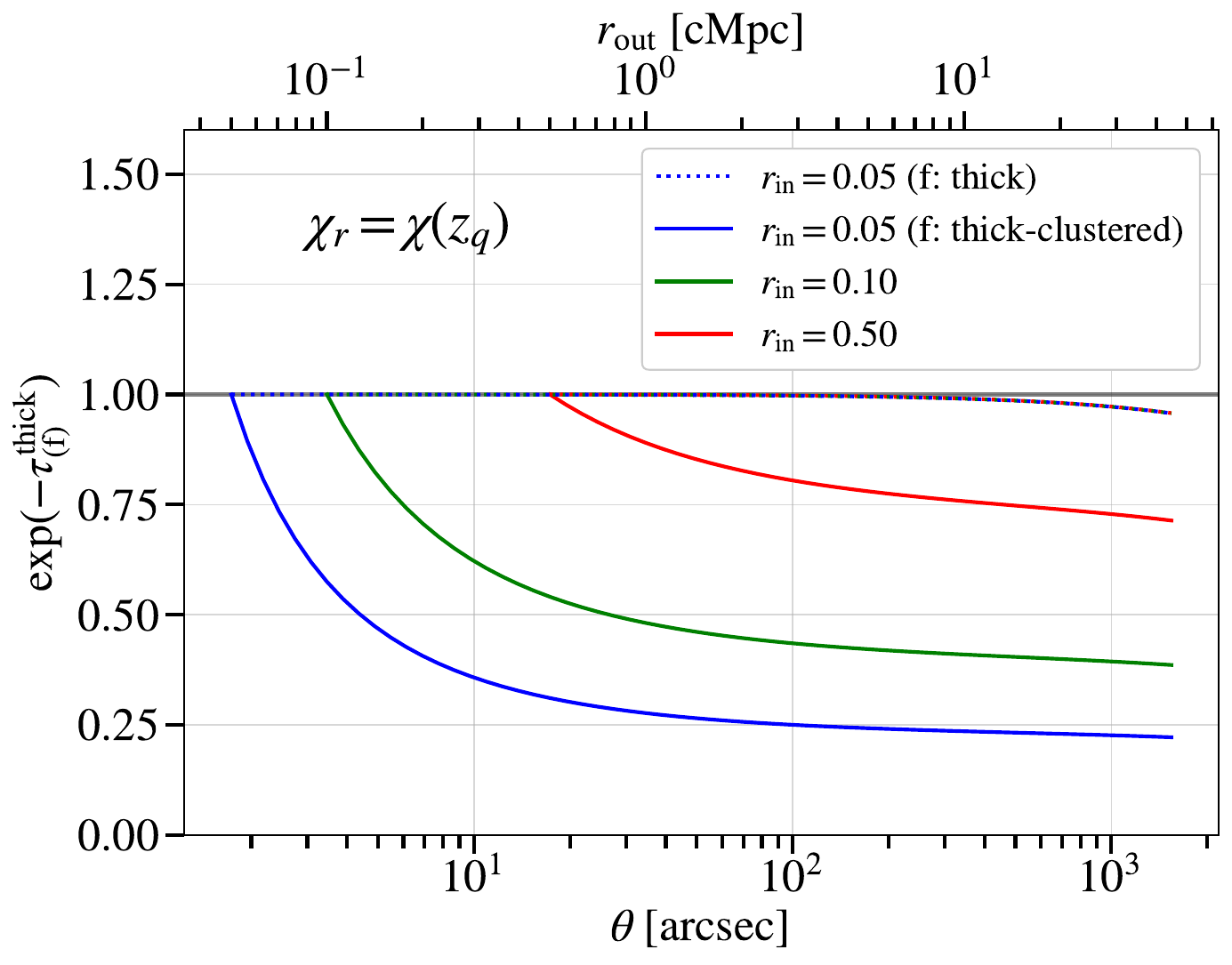}
\end{center}
\caption{\label{fig:att_f} Transmission of quasar ionizing photons. Each line corresponds to the transmission of ionizing photons emitted from a quasar at $z_{q}=2.5$, which are attenuated by optically thick gas clouds with the incidences of $l_{q}$ (solid lines) and $l_{\rm IGM}$ (dotted lines). The three colors represent various starting points at different distances from the quasar: $r_{\rm in} [{\rm cMpc}] =$ 0.05 (blue), 0.10 (green), and 0.50 (red). The lower horizontal axis is the line-of-sight angle, $\theta$, of the ionization front at the distance of $\chi_{r}=\chi(z_{q})$ and the upper horizontal axis is the comoving distance, $r_{\rm out}$ (i.e., $\chi_{qr}$), corresponding to $\theta$. 
}
\end{figure}
 
In Sections.~\ref{sec:scattering} and \ref{sec:fluorescence}, when calculating the Ly$\alpha$ emissivity at the point of $R$, we implicitly assumed that the quasar continuum photons are not attenuated by HI gas clouds on their way from the quasar to the point. However, the expression for the Ly$\alpha$ emissivity exactly means that quasar Ly$\alpha$ or ionizing photons are absorbed at any position by the IGM. Then, we estimate the attenuation of quasar ionizing photons due to LLSs, which are reprocessed to Ly$\alpha$ photons via fluorescence and turned to the most dominant contribution to the quasar-induced Ly$\alpha$ emission. The optical depth of ionizing photons propagating through LLSs (see equation~\ref{eq:j_lya_f_thick}) is
\begin{eqnarray}
    \tau_{(f)}^{\rm thick} &=& \sum_{\rm thick} \int_{s_{\rm in}}^{s_{\rm out}} n_{\rm c}\sigma_{\rm c} \mathrm{d}s
    \nonumber \\
    &\simeq& \frac{1}{1+z_{q}}\int_{r_{\rm in}}^{r_{\rm out}} l_{q}^{\rm thick}(z_{q}, r) \left({\rm or}\ l_{\rm IGM}^{\rm thick}(z_{q})\right) \mathrm{d}r, \label{eq:att_f_thick}
\end{eqnarray}
where $s_{\rm in}$ ($s_{\rm out}$) is the physical distance from the quasar to the point where ionization photons leave (reach) and $r_{\rm in}$ ($r_{\rm out}$) is its comoving distance. Fig.~\ref{fig:att_f} shows the transmission: $\exp(-\ \tau_{(f)}^{\rm thick})$ for LLSs with the incidences of $l_{q}$ (solid lines) and $l_{\rm IGM}$ (dotted lines). Each color corresponds to a different starting point: $r_{\rm in} [{\rm cMpc}] =$ 0.05 (blue), 0.10 (green), and 0.50 (red). We can see that the cases with $l_{\rm IGM}$ are almost perfectly overlapped with each other and then independent of $r_{\rm in}$. On the other hand, the cases with $l_{q}$, considering the clustering enhancement, strongly depend on $r_{\rm in}$ because the quasar-absorber cross-correlation rapidly increases in the inner region, reflecting the power-law nature. The distance $r_{\rm in}$ here corresponds to a radius from quasars, within which the typical quasar continuum spectrum (not attenuated by the intervening IGM) is determined. Here, for a simple estimation, we assume that $r_{\rm in}$ can be evaluated by a typical radius of the circumgalactic medium (CGM), which surrounds galaxies out to distance scales of $\sim10^2$ pkpc \citep[e.g.,][]{2017ARA&A..55..389T}. We can then guess that the transmission of ionizing photons from a quasar at $z_{q} =2.5$ is higher than the one with $r_{\rm in}=0.10$ cMpc (green), i.e., $\gsim40\%$ at a distance of 30 cMpc. We note that since the transmission is very sensitive to the power-law index $\gamma$, it is difficult to accurately predict the attenuation at this point. 

In addition, we also assumed that the quasar-induced Ly$\alpha$ photons are not attenuated by HI gas clouds on their way from the point of $R$ to the observer. The attenuation of Ly$\alpha$ photons due to HI atoms in the IGM can be estimated using the transmission: $\exp(-\tau_{\alpha, {\rm eff}})$, where $\tau_{\alpha, {\rm eff}}$ is the Ly$\alpha$ effective optical depth, and we found that it is roughly $80\%$ for $z_{r}=2.5$ ($\tau_{\alpha, {\rm eff}}=0.23$) and $40\%$ for $z_{r}=4.0$ ($\tau_{\alpha, {\rm eff}}=0.88$) \citep{2008ApJ...681..831F}.

The above discussion implies that the attenuation of quasar ionizing photons or reprocessed Ly$\alpha$ photons is insignificant, and they are only reduced to a fraction as long as we assume the typical size of proximity zone and the quasar redshift range of $z\lsim4$. However, can we capture the quasar-induced Ly$\alpha$ photons more effectively? A simple approach is to map Ly$\alpha$ emissions around quasars with more pixels. As we discussed in Section~\ref{sec:forecast}, increasing the number of pixels improves the SNR ratio of the quasar-Ly$\alpha$ emission cross-correlation. In this sense, a natural extension of our study is using integral field spectrographs (IFSs), e.g., the Palomar/Keck Cosmic Web Imager \citep[PCWI/KCWI,][]{2010SPIE.7735E..0PM,2018ApJ...864...93M} or the Multi Unit Spectroscopic Explorer \citep[MUSE,][]{2010SPIE.7735E..08B}. Previous studies of extended Ly$\alpha$ emissions around quasars with these IFSs, e.g., CWI \citep[][]{2014ApJ...786..106M,2019ApJS..245...23C} or MUSE \citep[][]{2016ApJ...831...39B,2019MNRAS.482.3162A,2019ApJ...887..196F,2019ApJ...880...47M,2021MNRAS.503.3044F}, targeted fields centered on a few dozen, sampled quasars. Therefore, it is worth comparing the gain from a larger number of pixels obtained around each quasar with IFS surveys, with the one from a larger sample of quasars achieved with fiber spectrograph surveys.

\begin{figure*}
\begin{tabular}{cc}
\begin{minipage}{0.5\hsize}
\begin{center}
\includegraphics[width=\linewidth]{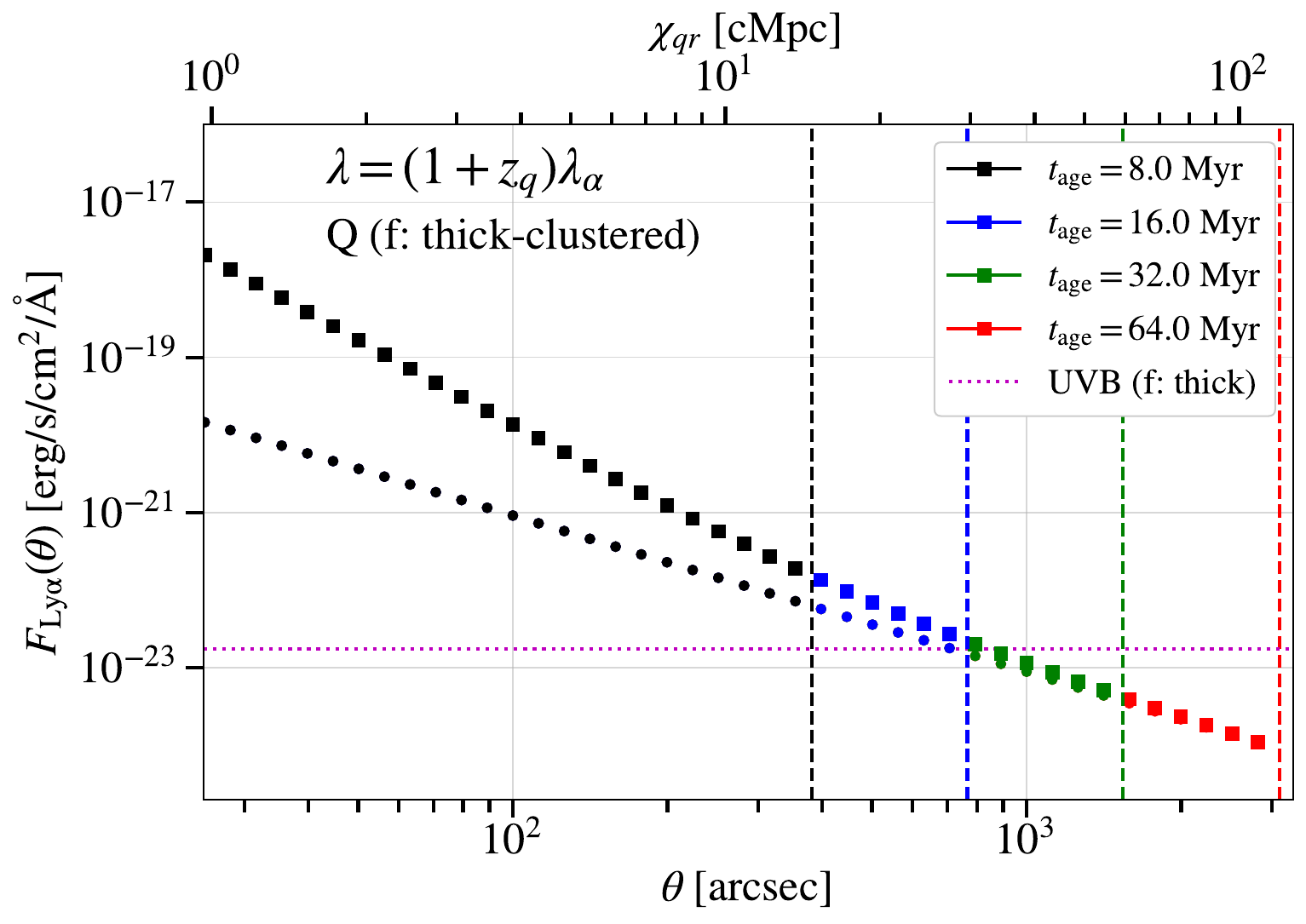}
\end{center}
\end{minipage} 
\begin{minipage}{0.5\hsize}
\begin{center}
\includegraphics[width=\linewidth]{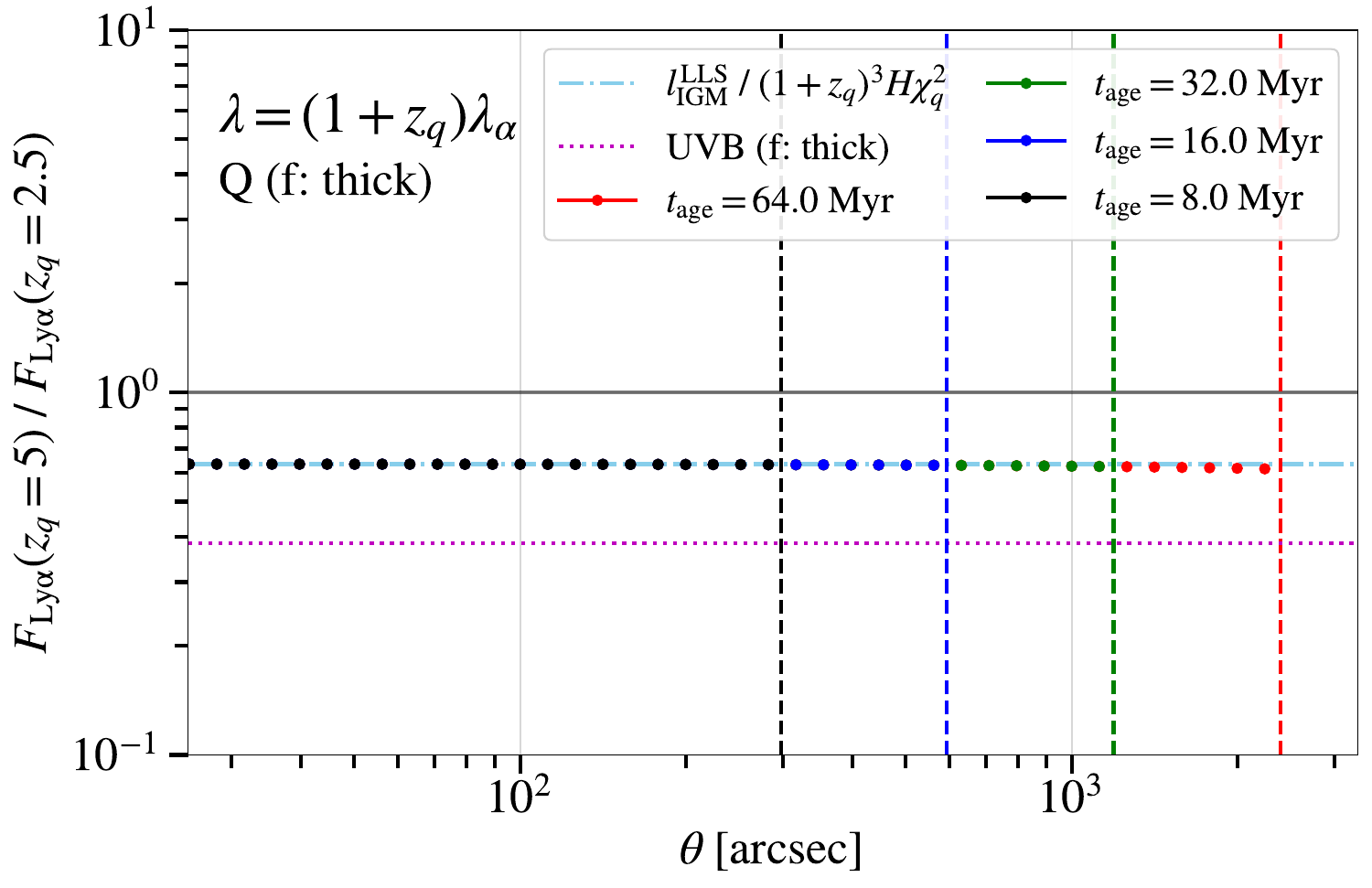}
\end{center}
\end{minipage}
\end{tabular}
  \caption{\label{fig:flux_s_the_comp} Redshift dependence of the fluorescence Ly$\alpha$ flux from LLSs. The left panel describes both cases with $l^{\rm LLS}_{q}$ (squares) and $l^{\rm LLS}_{\rm IGM}$ (dots) in the same manner as Fig.~\ref{fig:flux_the}, but at $z_{q}=5.0$. The right panel corresponds to the ratio between two different redshifts, $z_{q}=2.5$ and $5.0$, for the case with $l^{\rm LLS}_{\rm IGM}$. We also show the analytical predictions for the redshift dependence with the light blue, dot-dashed line.}
\end{figure*}

\subsection{Redshift dependence \label{sec:z_dependence}}

To study the effect of quasar illumination on the surrounding IGM, we considered quasars at a redshift of $z = 2.5$. Here, we evaluate the redshift dependence of the Ly$\alpha$ flux (per unit wavelength) for fluorescence emission from LLSs, which is the dominant contribution to the quasar-induced Ly$\alpha$ emission. As we discussed in Sections~\ref{sec:F_Q_f} and \ref{sec:F_lya}, in the optically thick case (to ionizing photons), each HI gas cloud within a cylinder of $A_{\rm f} \times \mathrm{d}s/\mathrm{d}\lambda$ reflects ionizing photons from a quasar, with the number flux of $\Phi_{q,{\rm LL}}$. From equations~(\ref{eq:j_lya_f_thick}), (\ref{eq:phi_q_ll}), and (\ref{eq:F_lya}), the redshift dependence without considering the clustering enhancement, i.e., the incidence of $l_{\rm IGM}$, is then described by
\begin{eqnarray}
\frac{l_{\rm IGM}^{\rm LLS}(z_q)\ \Phi_{q,{\rm LL}}}{(1+z_q)^{4}(H\lambda_{\alpha, {\rm obs}})} \propto \frac{l_{\rm IGM}^{\rm LLS}(z_q)}{(1+z_{q})^{3}H\chi_{q}^{2}}, \label{eq:z_depend}
\end{eqnarray}
where we used the fact that the quasar flux at a position of $(\theta, z_{r} = z_{q})$ roughly scales as $[\theta\chi_{q}/(1+z_{q})]^{-2}$. The left panel of Fig.~\ref{fig:flux_s_the_comp} shows the angular dependence of the fluorescence Ly$\alpha$ flux for a quasar at $z_{q} = 5.0$. There is no significant difference with the case of $z_{q} = 2.5$. We can also see that the redshift dependence is relatively weak from the right panel of Fig.~\ref{fig:flux_s_the_comp}, which shows the ratio of these two redshifts for the incidence of $l^{\rm LLS}_{\rm IGM}$ and its analytical prediction (equation~\ref{eq:z_depend}). Thus, while the incidence of absorbers per unit redshift, $\left(c/(1+z)H\right)l_{\rm IGM}$, monotonically increases with redshift \citep[e.g.,][]{2014MNRAS.442.1805I}, the gain is largely offset by other distance factors.

We comment on some concerns that we should be aware of when considering quasars at higher redshifts. The ionization rate due to the UV background, $\Gamma_{{\rm UVB},{\rm ion}}$, does not evolve in the redshift range of $z = 2{\rm -}5$ and remains a constant value of $\sim 10^{-12}\ \rm{s}^{-1}$. However, $\Gamma_{{\rm UVB},{\rm ion}}$ is an order of magnitude lower at $z \sim 6$ \citep[e.g.,][]{2013MNRAS.436.1023B}, which leads to an increase in the ionization timescale$\lsim \Gamma_{{\rm UVB},{\rm ion}}^{-1}$. Thus, at such high redshifts, the ionization front has a ``thickness'' due to a finite amount of the ionization time, which is still less than the typical quasar age. An important difference from the situation at low redshifts is that the collisional excitation rate can be higher than the recombination rate in the spherical shell of the ionization front due to the sufficient number of HI atoms \citep[e.g.,][]{2008ApJ...672...48C}. At $z\gsim6$, therefore, we need to consider the collisional excitation process as well as the resonant scattering or fluorescence process when modeling Ly$\alpha$ emission on the quasar light echo surface.

\section{Conclusions} \label{sec:conclusion}

In this study, we started with a Ly$\alpha$ emission intensity mapping method using galaxy spectra from wide-field spectroscopic surveys and considered the contribution of Ly$\alpha$ emission from the IGM due to quasar radiation, with the aim of inferring quasar properties, e.g., the quasar age, from the observed Ly$\alpha$ intensity map. To do this, we constructed a model that describes the quasar-induced Ly$\alpha$ emission by resonant scattering and fluorescence, taking into account the effect of the light cone, and estimated the quasar-induced contribution to the stacked spectrum of Ly$\alpha$ emission. Our achievements and findings in this paper are as follows:

\begin{enumerate}[leftmargin=0.6cm]
\item We adopted an analytic model for the neutral hydrogen column density distribution to estimate the incidences of the LAF and LLSs per physical length. In particular, the clustering enhancement in the dense environment around quasars was taken into account by a power-law function for the quasar-absorber cross-correlation (Fig.~\ref{fig:incidence_thick}). We then derived expressions for the Ly$\alpha$ volume emissivity induced by resonant scattering and fluorescence and for the Ly$\alpha$ flux per unit wavelength that is observed by spectroscopic fibers (Fig.~\ref{fig:around_R}). 

\item Then, the flux of the quasar-induced Ly$\alpha$ emission along a line of sight is given by a function of the line-of-sight separation from the quasar, and has a peak at the wavelength corresponding to the point closest to the quasar and a long wavelength cutoff at the light echo surface, depending on the quasar age (Fig.~\ref{fig:flux_all_theta}). The Ly$\alpha$ flux basically decreases with the separation angle, reflecting the distance decay of quasar radiation. Moreover, comparing different types of the quasar-induced Ly$\alpha$ emission (i.e., by scattering/fluorescence and from optically thin/thick gas clouds), we found that the most dominant contribution comes from the fluorescence of optically thick clouds, i.e., LLSs (Fig.~\ref{fig:flux_all_type}). 

\item We calculated the quasar-Ly$\alpha$ emission cross-correlation for the fluorescence due to LLSs (Fig.~\ref{fig:x_corr}) and its SNR for the PFS survey (Fig.~\ref{fig:SNR_PFS}), and then found that at distances of $\lsim20\ \rm{cMpc}\ \rm{h}^{-1}$, the SNR for quasars with ages older than 10~Myr is higher than $10^{-2}$. Furthermore, compared with a cross-correlation model for evaluating the contribution due to star-forming galaxies, the quasar-induced contribution can account for $\sim10\%$ of the model fitted to the BOSS and eBOSS data in the outer region of $\gsim10\ \rm{cMpc}\ \rm{h}^{-1}$. In conclusion, considering a large uncertainty in estimating the Ly$\alpha$ luminosity density, we found that our model for the quasar-induced Ly$\alpha$ emission, combined with the contribution of star-forming galaxies, is not in conflict with these previous measurements.

\item Possible contributions to the Ly$\alpha$ emission excess around quasars fall into two categories: those related to quasar illumination and those unrelated. In this paper, we have {\it not} considered the following:
    \begin{enumerate}[leftmargin=0.3cm]
    \item[] {\it Related to quasar illumination:}
    \item Ly$\alpha$ broad emission line 
    \item anisotropy of quasar radiation 
    \item[]
    \item[] {\it Unrelated to quasar illumination:}
    \item contribution of clustered external radiation sources, e.g., star-forming galaxies, to the IGM illumination
    \item collisional excitation (cooling radiation)
    \end{enumerate}

\item We also discussed the efficiency and the redshift dependence. While we used a typical quasar continuum spectrum to estimate the quasar ionizing flux, they are actually attenuated by LLSs on their way. Assuming that the quasar radiation is emitted at a typical radius of the CGM, the transmission of ionizing photons from a quasar at $z_{q} =2.5$ is $\gsim40\%$ at a distance of 30 cMpc (Fig.~\ref{fig:att_f}). Furthermore, we found that the fluorescence Ly$\alpha$ flux from LLSs does not depend much on the quasar redshift for the case without considering the clustering enhancement (Fig.~\ref{fig:flux_s_the_comp}). 
\end{enumerate}

In this paper, we have focused in particular on the quasar age as a quasar property. However, the Ly$\alpha$ emission induced by quasar radiation also depends on the quasar luminosity evolution. Ly$\alpha$ photons produced in the IGM at different distances from the observer (i.e., at different wavelengths) have traveled through different lengths, which means that they were emitted from the quasar at different times. Thus, the induced Ly$\alpha$ emissions at different wavelengths represent the history of the quasar activity, while the quasar spectrum reflects the luminosity only at the quasar redshift. For instance, in terms of the fluorescence Ly$\alpha$ flux along the spectrum stacked within each separation distance bin (Fig.~\ref{fig:flux_all_theta}), photons with the rightmost wavelength were emitted by the quasar at the moment it turned on (``oldest'' photons), and the shorter wavelengths correspond to photons that were emitted at later times (``younger'' photons). In this sense, a more realistic model, including potential contributions that we missed here, might allow us to constrain the quasar luminosity evolution over its lifetime rather than just its age. 

While the contribution of external radiation sources such as star-forming galaxies is not directly related to the properties of the quasar illumination, it does give us information about quasar clustering, since external sources, as well as quasars, trace the large-scale dark matter distribution \citep[e.g.,][]{2016MNRAS.457.3541C}. Therefore, a comparison of our scenario with an approach to determine quasar ages from quasar clustering measurements \citep{2001ApJ...547...12M} would provide more stringent constraints on quasar physics or environments. Although we expected to stack galaxy fibers around quasars of the same age to improve the SNR, sampling quasars by their luminosity or host halo mass allows us to learn the relationship between quasar age and other properties.

In addition to ongoing survey projects, e.g., PFS or DESI, several spectroscopic surveys will be launched in the next 15 years \citep[][]{2022arXiv220307506F}. In particular, the Maunakea Spectroscopic Explorer \citep[MSE,][]{2019arXiv190404907T} and MegaMapper \citep{2019BAAS...51g.229S} are expected to observe galaxies with higher number densities (by about an order of magnitude) and over wider redshift ranges ($1.6<z<4$ and $2<z<5$, respectively), and in addition SpecTel \citep{2019BAAS...51g..45E} may reach an even higher number density (by factors of magnitude) \citep[][]{2021JCAP...12..049S}. 
While we still need to consider how to optimize our estimator to efficiently constrain parameters from the measured Ly$\alpha$ intensity, not limited to the quasar-Ly$\alpha$ emission cross-correlation, these prospects are quite encouraging in that we can significantly improve the SNR.

\section*{Acknowledgements}

We would like to thank J. Xavier Prochaska for useful and stimulating discussions during the early stages of this work. RH is grateful to Daniel J. Eisenstein and Koki Kakiichi for helpful discussions. We also thank David H. Weinberg, Paul Martini, and Zheng Zheng for useful comments on an earlier draft and the revision. This work made use of {\tt Astropy},\footnote{\url{http://www.astropy.org}} a community-developed core Python package and an ecosystem of tools and resources for astronomy \citep{2013A&A...558A..33A,2018AJ....156..123A,2022ApJ...935..167A}, and {\tt CAMB},\footnote{\url{https://camb.info}} a cosmology code for calculating cosmological observables \citep{2000ApJ...538..473L}. This work was supported in part by World Premier International Research Center Initiative (WPI Initiative), Ministry of Education, Culture, Sports, Science and Technology, Japan, and Japan Society for the Promotion of Science KAKENHI Grant Numbers JP24H00215, JP23H00131, JP20H05850, JP20H05855, JP19H00677, and JP19J00513. RH acknowledges support from JSPS Research Fellowships for Young Scientists and JSPS Overseas Research Fellowships.

\section*{Data Availability}

The code used in this article and the resulting data shown in the figures will be shared on reasonable request to the corresponding author.



\bibliographystyle{mnras}
\bibliography{ms}



\begin{onecolumn}

\appendix

\section{HI column density distribution} \label{app:f_N_HI}

In Section~\ref{sec:f_N_HI}, we introduced a distribution function of HI column density $N_{\rm HI}$ at a redshift $z$, 
\begin{eqnarray}
    \frac{\partial^2 n}{\partial z \partial N_{\rm HI}}
        = f_{\rm LAF}(z)g_{\rm LAF}(N_{\rm HI}) + f_{\rm DLA}(z)g_{\rm DLA}(N_{\rm HI}),  
\end{eqnarray}
which was presented by \citet{2014MNRAS.442.1805I}. Here, we give the explicit expressions for each term and the model parameters assumed in our study, which are given in Table~1 of the same paper.

The total contribution from all absorbers to consists of two components, LAF and DLA, which were named after the Ly$\alpha$ forest (LAF) and damped Ly$\alpha$ systems (DLAs), respectively, to reflect their dominant contribution. The dependence on $N_{\rm HI}$ is described as 
\begin{eqnarray}
    g_{i}(N_{\rm HI})
        = B_{i}N_{\rm HI}^{-\beta_{i}}\mathrm{e}^{-N_{\rm HI}/N_{\rm c}},
\end{eqnarray}
where the subscript $i = \{{\rm LAF, DLA}\}$, $\beta_{i}$ is the power-law index for each component, $B_{i}$ is the normalization defined by boundaries $N_{\rm l}$ and $N_{\rm u}$ (i.e., $\int^{N_{\rm u}}_{N_{\rm l}}g_{i}(N_{\rm HI})\mathrm{d}N_{\rm HI}=1$), and $N_{\rm c}$ is the cut-off column density. The adapted values are as follows: $\beta_{\rm LAF} = 1.7, \beta_{\rm DLA} =0.9, N_{\rm l}=10^{12}\ {\rm cm^{-2}}, N_{\rm u}=10^{23}\ {\rm cm^{-2}}, N_{\rm c}= 10^{21}\ {\rm cm^{-2}}$. The redshift evolution function for each component is given by
\begin{eqnarray}
    f_{\rm LAF}(z) &=& \mathcal{A}_{\rm LAF}
\begin{cases}
    \left(\frac{1+z}{1+z_{\rm LAF, 1}}\right)^{\gamma_{\rm LAF, 1}} &\quad (z < z_{\rm LAF, 1})  \\[1em]
    \left(\frac{1+z}{1+z_{\rm LAF, 1}}\right)^{\gamma_{\rm LAF, 2}} &\quad (z_{\rm LAF, 1} \leq z < z_{\rm LAF, 2})  \\[1em]
    \left(\frac{1+z_{\rm LAF, 2}}{1+z_{\rm LAF, 1}}\right)^{\gamma_{\rm LAF, 2}} \left(\frac{1+z}{1+z_{\rm LAF, 2}}\right)^{\gamma_{\rm LAF, 3}} &\quad (z_{\rm LAF, 2} \leq z) 
\end{cases}
, \\[2em]
    f_{\rm DLA}(z) &=& \mathcal{A}_{\rm DLA}
\begin{cases}
    \left(\frac{1+z}{1+z_{\rm DLA, 1}}\right)^{\gamma_{\rm DLA, 1}} &\quad (z < z_{\rm DLA, 1})  \\[1em]
    \left(\frac{1+z}{1+z_{\rm DLA, 1}}\right)^{\gamma_{\rm DLA, 2}} &\quad (z_{\rm DLA, 1} \leq z) 
\end{cases}
,
\end{eqnarray}
with the model parameters: $\mathcal{A}_{\rm LAF}=500, z_{\rm LAF, 1\mathchar`-2}=\{1.2, 4.7\}, \gamma_{\rm LAF, 1\mathchar`-3}=\{0.2, 2.7, 4.5\}, \mathcal{A}_{\rm DLA}=1.1, z_{\rm DLA, 1}=\{2.0\}, \gamma_{\rm DLA, 1\mathchar`-2}=\{1.0, 2.0\}$.

\end{onecolumn}


\bsp	
\label{lastpage}
\end{document}